\documentclass[twocolumn,letterpaper]{aa520}
\usepackage[dvips]{graphicx}
\usepackage{parskip}
\usepackage{rotating}
\usepackage{txfonts}

\begin{document}

\title{The young B-star quintuple system HD~155448\thanks
      {Based on observations collected at the European Southern
       Obser\-vatory (programme IDs 075.C-0091(B) and 075.C-0091(C),
       together with archival and technical data)}}

\author{O.~Sch\"utz
        \inst{1}
        \and
        G.~Meeus
        \inst{2}
        \and
        A.~Carmona
        \inst{3, 4}
        \and
        A.~Juh\'asz
        \inst{5}
        \and
        M. F.~Sterzik
        \inst{1}
        }

\offprints{oschuetz@gmail.com}

\institute{European Southern Observatory, Alonso de Cordova 3107,
           Santiago 19, Chile
           \and
           Universidad Aut\'onoma de Madrid, Departamento de F\'isica 
           Te\'orica C-XV, 28049 Madrid, Spain
           \and
           ISDC Data Centre for Astrophysics, University of Geneva, 
           chemin d'Ecogia 16, 1290 Versoix, Switzerland
           \and 
           Observatoire de Gen\`eve, University of Geneva, 
           chemin des Maillettes 51, 1290 Sauverny, Switzerland 
           \and
           Leiden Observatory, Leiden University, P.O.\ Box 9513, 
           NL-2300 RA Leiden, The Netherlands
           }

\date{Received / Accepted }

\abstract{

Until now, HD~155448 has been known as a post-AGB star and listed as a 
quadruple system. In this paper, we study the system in depth and reveal 
that the B component itself is a binary and that the five stars HD~155448~A, 
B1, B2, C, and D likely form a comoving stellar system. From a 
spectroscopic analysis we derive the spectral types and find that all 
components are B dwarfs (A: B1V, B1: B6V, B2: B9V, C: B4Ve, D: B8V). Their 
stellar ages put them close to the ZAMS, and their distance is estimated to 
be $\sim$2~kpc. Of particular interest is the C component, which shows strong 
hydrogen and forbidden emission lines at optical wavelengths. All emission 
lines are spatially extended in the eastern direction and appear to have a similar
velocity shift, except for the [\ion{O}{i}] line. In the IR images, we see an 
arc-like shape to the 
northeast of HD~155448~C. From the optical up to 10~$\mu$m, most circumstellar 
emission is located at distances between $\sim$1.0\arcsec\ and 3.0\arcsec\ from 
HD~155448~C, while in the $Q$ band the arc-like structure appears to be in 
contact with HD~155448~C. 
The Spitzer and VLT/VISIR mid-IR spectra show that the circumstellar material 
closest to the star consists of silicates, while polycyclic aromatic 
hydrocarbons (PAH) dominate the emission at distances $>$1\arcsec\, with bands 
at 8.6, 11.3, and 12.7~$\mu$m. 
We consider several scenarios to explain the unusual, asymmetric, arc-shaped 
geometry of the circumstellar matter. The most likely explanation is an 
outflow colliding with remnant matter from the star formation process.


         }

\maketitle

\section{Introduction}
\label{sect:introduction}

\object{\mbox{HD 155448}} (alias \mbox{SAO 208540}, or \mbox{HIP 84228}) 
was described as a B9 object in the ``Catalogue of two-dimensional spectral 
types for the HD stars, Vol. 3'' (Houk \cite{Houk}). Three other members of 
this system are known according to the ``Catalogue of the components of 
double and multiple stars (CCDM)'' (Dommanget \& Nys \cite{Dommanget}). The 
system's Hipparcos parallax of 1.65~mas (Perryman et al.\ \cite{Perryman}) 
would correspond to a distance of roughly $\mathrm{d} = 600$~pc, but the 
uncertainty of the parallax is large (1.99~mas).

Up to now, all studies have described this system as old and evolved. In their 
sample of 42 IRAS point source objects, Van der Veen et al.\ (\cite{Veen}) 
considered IRAS~17097$-$3210 (alias \object{\mbox{HD 155448}}) -- based on 
its IRAS colours -- as a transition object from the AGB to the planetary 
nebula (PN) stage. These authors also speculated about PAH emission at 
7.8~$\mu$m. However, in a survey of 2703 IRAS point sources at Dwingeloo, 
Effelsberg, and Parkes, examining the 1612~MHz OH-transition -- which is one
of the strongest lines for AGB stars, but decreases towards the PN stage 
-- Te Lintel Hekkert et al.\ (\cite{TeLintel}) did not detect emission from 
\object{\mbox{HD 155448}}.

Malfait et al.\ (\cite{Malfait}) points out that, in order to distinguish  
between Herbig Ae/Be stars and post-AGB stars, one should also consider the 
([$B-V$],[$U-B$])-colour diagram, since post-AGB stars show an important Balmer 
discontinuity, while the regions populated by both classes overlap in 
infrared colour-colour diagrams. They also show that several other 
currently well-accepted Herbig Ae/Be stars had previously been classified 
as transition objects between the AGB and the planetary-nebula stage. From 
the absence of a prominent near-IR excess (the excess only starts beyond 
4.5~$\mu$m), these authors consider \object{\mbox{HD 155448}} to be a 
Vega-type candidate. Spectral energy distributions for the integrated fluxes 
have been given by Van der Veen et al.\ (\cite{Veen}) and Malfait et 
al.\ (\cite{Malfait}).

\begin{table*}[t]
  \centering
  \caption{Summary of all imaging data presented in this work.}
  \setlength\tabcolsep{18.3pt}
  \begin{tabular}{lccccr}
    \hline
    \hline
    \noalign{\smallskip}
    Passband                 &  Instrument           &  Date  
                             &  Pixel scale          
                             &  $t_{\rm exp}$          &  Airmass \\
                             &                       &  of night's begin
                             &  [$''$]               
                             &  [sec]                &          \\
    \noalign{\smallskip}
    \hline
    \noalign{\medskip}
    $B$                      &  EFOSC2               &  Feb 26, 2006
                             &  0.157                
                             &  8                    &  1.3 
                             \smallskip \\
    $V$                      &  EFOSC2               &  Feb 26, 2006
                             &  0.157                
                             &  8                    &  1.2 
                             \smallskip \\
    $R$                      &  EFOSC2               &  Feb 26, 2006
                             &  0.157                
                             &  8                    &  1.2 
                             \smallskip \\
    \hline
    \noalign{\medskip}
    $J$                      &  SOFI                 &  May 07, 2004
                             &  0.144                
                             &  100                  &  1.0         
                             \smallskip \\
    $H$                      &  ADONIS               &  Jun 08, 2000
                             &  0.100                
                             &  600                  &  1.0         
                             \smallskip \\
    $H$                      &  SOFI                 &  May 07, 2004
                             &  0.144                
                             &  100                  &  1.0         
                             \smallskip \\
    NB 1.64 $\mu$m           &  NACO                 &  Sep 09, 2005
                             &  0.027                
                             &  420                  &  1.1--1.5  
                             \smallskip \\
    $SK$                     &  ADONIS               &  Jun 08, 2000
                             &  0.100                
                             &  600                  &  1.0         
                             \smallskip \\
    $Ks$                     &  SOFI                 &  May 07, 2004
                             &  0.144                
                             &  100                  &  1.0         
                             \smallskip \\
    NB 2.12 $\mu$m           &  NACO                 &  Sep 09, 2005
                             &  0.027                
                             &  435                  &  1.1--1.5  
                             \smallskip \\
    \hline
    \noalign{\medskip}
    PAH2 \quad \enspace \, (11.25~$\mu$m)      
                             &  VISIR                &  Apr 27 -- Aug 05, 2005
                             &  0.075                
                             &  5425                 &  1.0--1.5  
                             \smallskip \\
    PAH2\_ref \enspace (11.88~$\mu$m) 
                             &  VISIR                &  Apr 27 -- Aug 05, 2005
                             &  0.075                
                             &  5425                 &  1.0--1.6  
                             \smallskip \\
    $Q2$ \quad \quad \quad\,(18.72~$\mu$m)       
                             &  VISIR                &  Aug 26 -- Sep 01, 2005
                             &  0.075                
                             &  7948                 &  1.0--1.2  
                             \smallskip \\
    \hline
    \noalign{\smallskip}
  \end{tabular}
  \label{table:ima-obs}
  \bigskip
  \medskip
\end{table*}

\begin{table*}[t]
  \centering
  \caption{Summary of all spectroscopic data presented in this work. 
           See Fig.\,\ref{fig:slits} and 
           Appendix~\ref{sect:obs-efosc-spec} for the definition of the 
           EFOSC2 slit orientations.}
  \setlength\tabcolsep{9.7pt}
  \begin{tabular}{lcccccr}
    \hline
    \hline
    \noalign{\smallskip}
    Passband                 &  Instrument           &  Date  
                             &  Slit                 &  Spec.\ res.
                             &  $t_{\rm exp}$          &  Airmass \\
                             &                       &  of night's begin
                             &  orientation          &  $R$
                             &  [sec]                &          \\
    \noalign{\smallskip}
    \hline
    \noalign{\medskip}
    Grism 5  \enspace \enspace (5200--9350\,\AA)
                             &  EFOSC2               &  Aug 19, 2007
                             &  East-West            &  $\sim$300
                             &  1700                 &  1.3 
                             \smallskip \\
    Grism 11 \enspace (3380--7520\,\AA)                
                             &  EFOSC2               &  Mar 01, 2006
                             &  A-C, B-C, C-D        &  $\sim$400
                             &  240                  &  1.2 
                             \smallskip \\
    Grism 11 \enspace (3380--7520\,\AA) 
                             &  EFOSC2               &  Apr 20, 2006
                             &  East-West            &  $\sim$400
                             &  2700                 &  1.0 
                             \smallskip \\
    Grism 18 \enspace (4700--6770\,\AA)
                             &  EFOSC2               &  Aug 19, 2007
                             &  East-West            &  $\sim$600
                             &  2000                 &  1.1 
                             \smallskip \\
    Grism 19 \enspace (4441--5114\,\AA)
                             &  EFOSC2               &  Aug 12, 2008
                             &  A-C, B-C, C-D        &  $\sim$3000
                             &  420--1800            &  1.0--2.0  
                             \smallskip \\
    Grism 20 \enspace (6047--7147\,\AA)
                             &  EFOSC2               &  Aug 12, 2008
                             &  A-C, East-West       &  $\sim$2500
                             &  420--1800            &  1.3--1.8  
                             \smallskip \\
    \hline
    \noalign{\medskip}
    8--13~$\mu$m      
                             &  VISIR                &  Apr 15 -- Sep 03, 2005
                             &  East-West            &  $\sim$350
                             &  1800                 &  1.0--1.4  
                             \smallskip \\
    SL \quad \quad \quad \,(5.5--14~$\mu$m)      
                             &  Spitzer/IRS          &  March 22, 2005
                             &  see \,Fig.\,\ref{fig:irs-slits}
                                                     &  $\sim$160
                             &  12                   &  n/a
                             \smallskip \\
    SH \quad \quad \quad (10--20~$\mu$m)       
                             &  Spitzer/IRS          &  March 22, 2005
                             &  see \,Fig.\,\ref{fig:irs-slits} 
                                                     &  $\sim$160
                             &  60                   &  n/a 
                             \smallskip \\
    LH \quad \quad \quad (19--38~$\mu$m)       
                             &  Spitzer/IRS          &  March 22, 2005
                             &  see \,Fig.\,\ref{fig:irs-slits} 
                                                     &  $\sim$160
                             &  24                   &  n/a 
                             \smallskip \\
    \noalign{\smallskip}
    \hline
  \end{tabular}
  \label{table:spec-obs}
  \bigskip
  \medskip
\end{table*}

Several important issues remain unclear:

\begin{itemize}

\item{
What is the evolutionary status of this system? Does the excess emission 
point towards a pre-main sequence stage in which the star is surrounded
by a disk? What is the composition of the material causing the IR excess 
emission?
}

\item{
Up to now it has not been proven whether the components are physically 
related. Are there more members to be discovered?
}

\item{
What is the actual distance of this system?
}

\end{itemize}

In this paper, we present the first high-resolution, multi-wavelength 
photometry and spectroscopy of the \object{HD 155448} system intended to
address the above questions. 
In Sect.~\ref{sect:starprops} we derive new spectral types for all 
components and constrain the system's evolutionary status and distance. 
In Sect.~\ref{sect:IRanalysis} we show that the B component itself is a 
binary, uncover arc-shaped IR emission near the C component, and derive 
the composition of the material emitting in the IR. After a discussion in 
Sect.~\ref{sect:discussion}, we round off with conclusions in 
Sect.~\ref{sect:conclusions}.

\begin{figure}[t]
  \centering
  \includegraphics[scale=.41, angle=0]{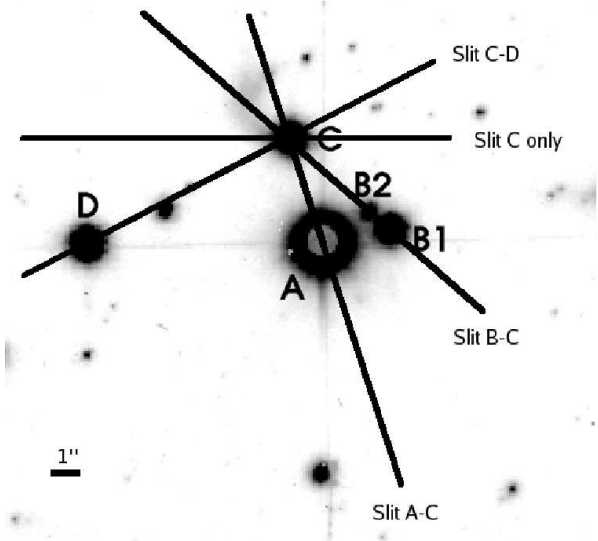}
  \caption{ADONIS $H$ band image showing the location of the five 
           components in the HD~155448 system. For the first time the B 
           companion is resolved into two sources, B1 and B2. On top of 
           this, we indicate the orientation of the EFOSC2 slits. {\sl Image 
           orientation here and in all following images is north up and
           east left.}}
  \label{fig:slits}
  \bigskip
  \medskip
\end{figure}

Tables~\ref{table:ima-obs} and~\ref{table:spec-obs} summarise the imaging
and spectroscopic data presented in this work. A detailed description of the 
observations and data reduction is given in Appendix~\ref{sect:observation}.
Tables~\ref{table:single-flux-1} and~\ref{table:single-flux-2} show the 
derived aperture photometry. The
slit orientation of the spectroscopic data is illustrated in 
Fig.\,\ref{fig:slits} for the optical spectra and in Fig.\,\ref{fig:irs-slits}
for the infrared data.

\section{Stellar properties derived from optical data}
\label{sect:starprops}

\subsection{HD 155448 system members}

\begin{figure}[t]
  \centering
  \includegraphics[scale=.46, angle=0]{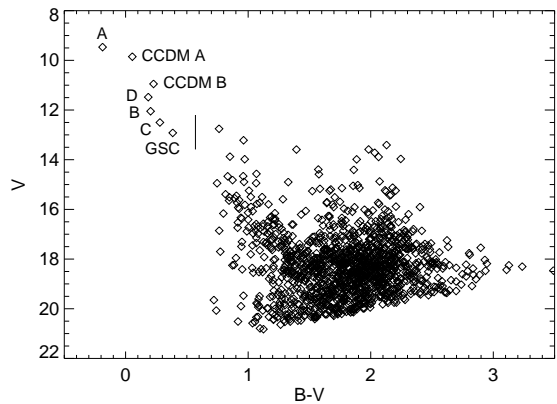}
  \caption{Colour-magnitude diagram of all $\sim$2000 objects in the 
           5.5\arcmin $\times$\,5.5\arcmin\ EFOSC2 field. The vertical line 
           in the top left quadrant separates 
           stars at the distance of HD~155448 from background objects. 
           Letters ABCD denote HD~155448 components (see in the text which 
           assumption was made for the C component). CCDM represents the 
           binary star \object{CCDM 
           J17130$-$3216AB}. GSC denotes \object{GSC-2 S22231309}. Sources 
           with B~$>$~21~mag are not displayed.}
  \label{fig:col-mag}
  \medskip
\end{figure}

In Table~\ref{table:single-flux-1}, we give $BVR$ photometry of the individual
components, apart from the B1 and B2 binary that was not resolved at optical 
wavelengths. 
%
Although \object{HD 155448} was often referred to as a quadruple system, 
there is no clear proof of the membership of all components to this system. 
Could there be even more components among the neighbouring stars in the 
EFOSC2 field? Therefore, we plot all objects in a colour-magnitude diagram 
($V$ vs.\ $B-V$; see Fig.\,\ref{fig:col-mag}). We lack $U$ band data for an 
explicit extinction correction, so we did not deredden the data, but the 
resulting diagram is clearly different from an HRD, thus demonstrating that 
these stars are not at similar distances. A group of seven objects is 
separated from the broader distribution of field stars and appears to be at 
a similar distance. Four of these stars are HD~155448 A, B, C, and D. To 
account for the additional reddening of HD~155448~C caused by the CS 
material, we dereddened this component using $A_V$ = 3.7, a value chosen
such that the remaining $A_V$ = 1.5 matches the mean reddening of the other
components A, B1, B2 
and D (see Table~\ref{table:spec-class}). We derived the corresponding $A_B$ 
of HD~155448~C according to the interstellar extinction curve given in 
Savage \& Mathis (\cite{Savage}). The remaining three stars of the subgroup 
shown in the upper left of Fig.\,\ref{fig:col-mag} are located 1--2\arcmin\ 
away from \object{HD 155448}. 
Although they could be at a similar distance, they are separated too far 
to be part of the HD~155448 system (see also Sect.\,\ref{distances} for a 
discussion of the distance of the \object{HD 155448} system). While we can 
exclude more system members by this test, the membership of HD~155448~C 
is not yet proven, although very likely: it is very unlikely to find five 
B-type stars of luminosity class V on such a small scale ($\sim$60 arcsec$^2$) 
by coincidence.

\begin{figure}[t]
  \centering
  \includegraphics[scale=.60, angle=0]{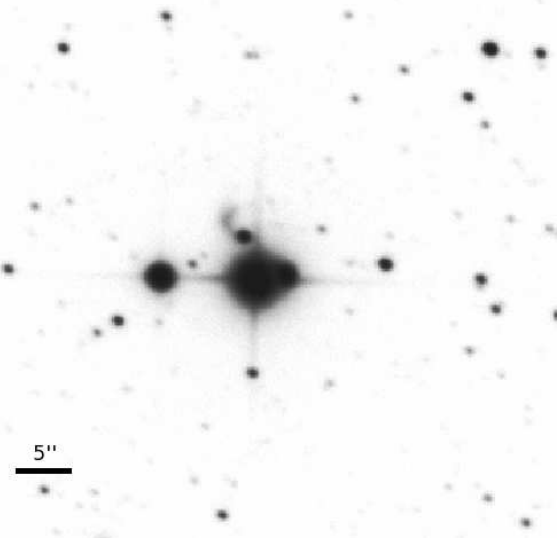}
  \caption{EFOSC2 $R$ band image. The cut levels were adjusted to also display
           the arc northeast of the C component, which may give the wrong 
           impression that components A and B were not resolved. See
           Fig.\,\ref{fig:slits} for an identification of the components.}
  \label{fig:efosc}
  \bigskip
  \medskip
\end{figure}

To estimate the likelihood of a chance grouping we considered galactic models.
However, since the \mbox{\object{HD 155448}} system is located in the direction 
towards the galactic centre (l$\sim$353, b$\sim$+04), we cannot apply common
galactic models. Alternatively, we investigated the average distribution of 
B0V\,--\,B9V stars in this region in the SIMBAD database, and find two B-type 
main sequence stars per deg$^2$ averaged over a 5\,$\times$\,5 deg$^2$ field, 
centred on the position of \mbox{\object{HD 155448}}.
However, this census is incomplete due to the uncertain limiting magnitude in 
the SIMBAD database. More than 90\% of the B-type stars found in this 
5\,$\times$\,5 
deg$^2$ field are from the HD catalogue, which according to Flynn \& Freeman 
(\cite{Flynn}) is complete down to a magnitude $V$\,=\,9.2 mag (but some 
objects in our SIMBAD test field are as faint as $V$\,=\,11.5 mag). The number 
of those B-type main sequence stars in the test field brighter than 
$V$\,=\,9.2 mag is one per deg$^2$. We want to know how many B stars we can 
expect down to $V$\,=\,12.2 mag, corresponding to the faintest HD~155448 
member, if we do not consider component C because of its extra extinction. In 
a standard assumption the number of stars multiplies on average by a factor
$\sim$2.5 for each magnitude (see also, e.g., Table~3 in Roach \cite{Roach}).
Based on the count of one B-type star brighter 
than $V$\,=\,9.2 mag per deg$^2$, this would result in about 16 B-type stars 
with luminosity class V brighter than $V$\,=\,12.2 mag that could be 
expected per deg$^2$. The HD~155448 components cover an area of 
$\sim$60~arcsec$^2$. Therefore, the probability that the entire HD~155448 
system is a chance alignment is $\leq$\,10$^{-5}$. We return to a discussion 
of the membership status in Sect.\,\ref{comotion}.

\subsection{Derivation of spectral types}

To derive the spectral type and luminosity class of the \object{HD 155448} 
candidate members, we carefully compared our EFOSC2 spectra with 
high-resolution (R$\sim$80000) spectra from the UVES Paranal Observatory 
Project spectral library\footnote{http://www.sc.eso.org/santiago/uvespop/} 
(Bagnulo et al.\ \cite{Bagnulo}), the ELODIE (R$\sim$42000) spectral 
library\footnote{http://atlas.obs-hp.fr/elodie/} (Soubiran et 
al.\ \cite{Soubiran}), and the theoretical high-resolution (R$\sim$500000) 
spectral library 
BLUERED\footnote{http://www.inaoep.mx/$\sim$modelos/bluered/bluered.html} by 
Bertone et al.\ (\cite{Bertone}). We proceed in a similar way to
Carmona et al.\ (\cite{Carmona}). First, the template spectra were 
degraded down to the EFOSC2 resolution. We then used a dedicated interactive 
IDL-based software\footnote{Available from A. Carmona: 
andres.carmona@unige.ch} that allowed us to compare the normalised target 
spectrum simultaneously (i.e., over-plotted) with a normalised template 
spectrum at a user-selected spectral feature. Target and template spectra 
were normalised either by the median or by a polynomial fit of second order 
to the flux in the windows used for the spectral comparison.

\begin{figure}[t]
  \centering
  \includegraphics[scale=.46, angle=0]{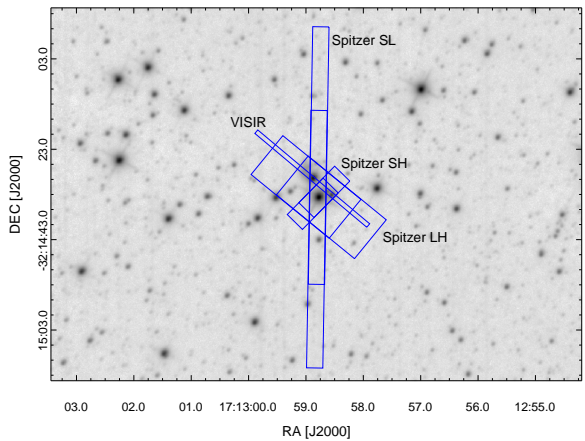}
  \caption{VISIR and {\it Spitzer} IRS slits and their orientation, plotted 
           on top of a SOFI $Ks$ band image.}
  \label{fig:irs-slits}
  \bigskip
\end{figure}

We compared our EFOSC2 spectra with the spectral templates spectral feature 
by spectral feature, employing a window of 20--100~\AA\ width around the 
central wavelength of each feature. The selected spectral features used for 
comparison are taken from the ``Atlas of stellar spectra'' by  Ginestet 
et al.\ (\cite{Ginestet}) and the features recommended by Morgan, Keenan, \& 
Kellman (\cite{Morgan}) from ``An atlas of stellar spectra with an 
outline of spectral classification''. To determine the luminosity class, we 
mainly employed the diagnostics of Ginestet et al.\ (\cite{Ginestet}),
described in the plates that show the effect of the luminosity for each 
sub-spectral type. In summary, the procedure consisted in the use of the 
Ginestet et al.\ (\cite{Ginestet}) and Morgan, Keenan, \& Kellman 
(\cite{Morgan}) works to guide the choice of which spectral diagnostics to 
analyse, and to use the UVES, ELODIE, or BLUERED spectral library templates 
to compare the strengths and shapes of the lines to our spectra. Visually it 
is relatively straightforward to determine which spectral template matches 
the spectrum under analysis best. However, in cases of degeneracy 
(especially due to rotation), the template spectrum that exhibited the 
smallest residuals (square root of the summed square of the difference 
between the normalised template and target spectrum) was adopted as best 
match of the target spectrum.

\begin{figure*}[t!]
  \centering
  \includegraphics[scale=.52, angle=0]{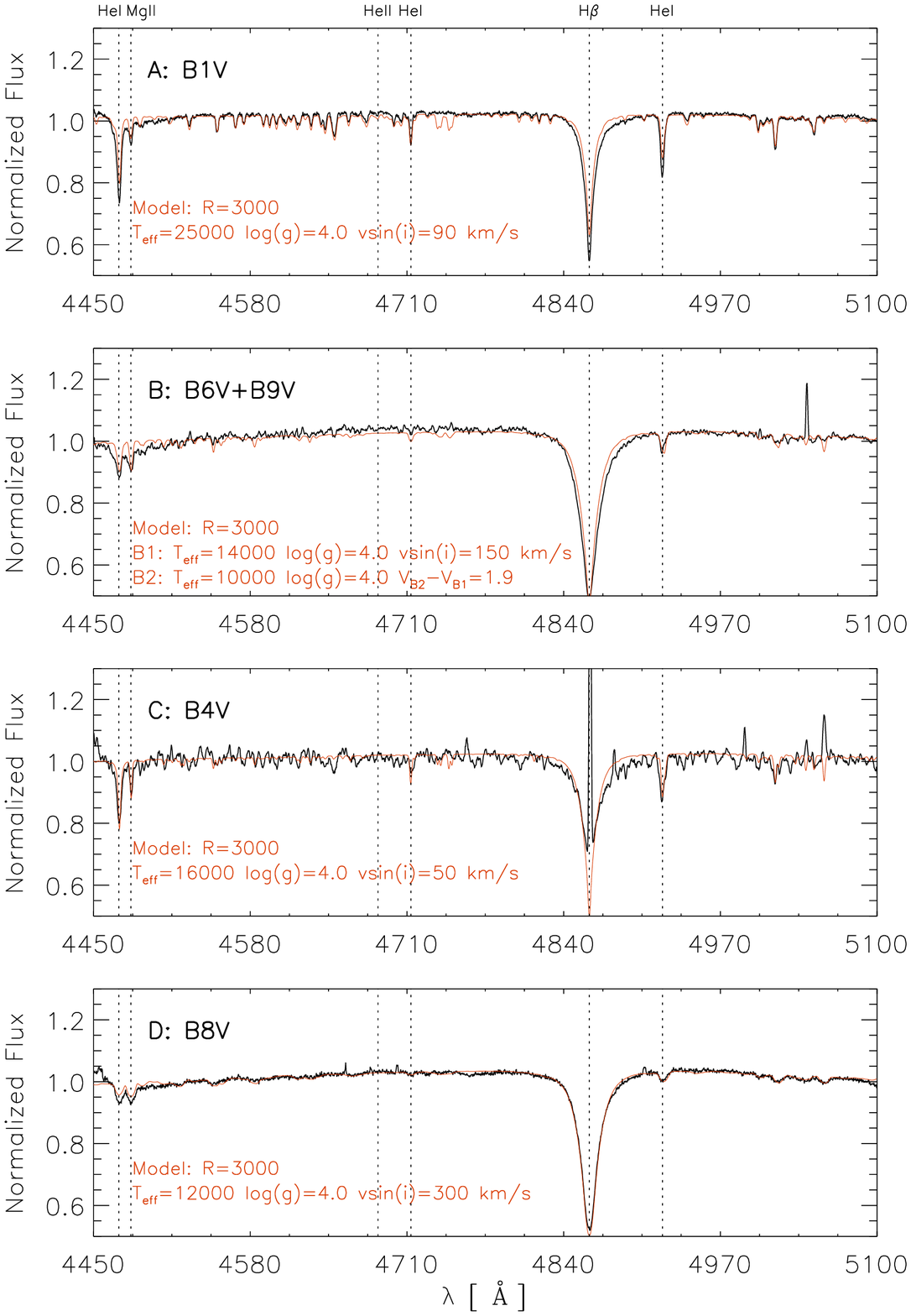}
  \includegraphics[scale=.52, angle=0]{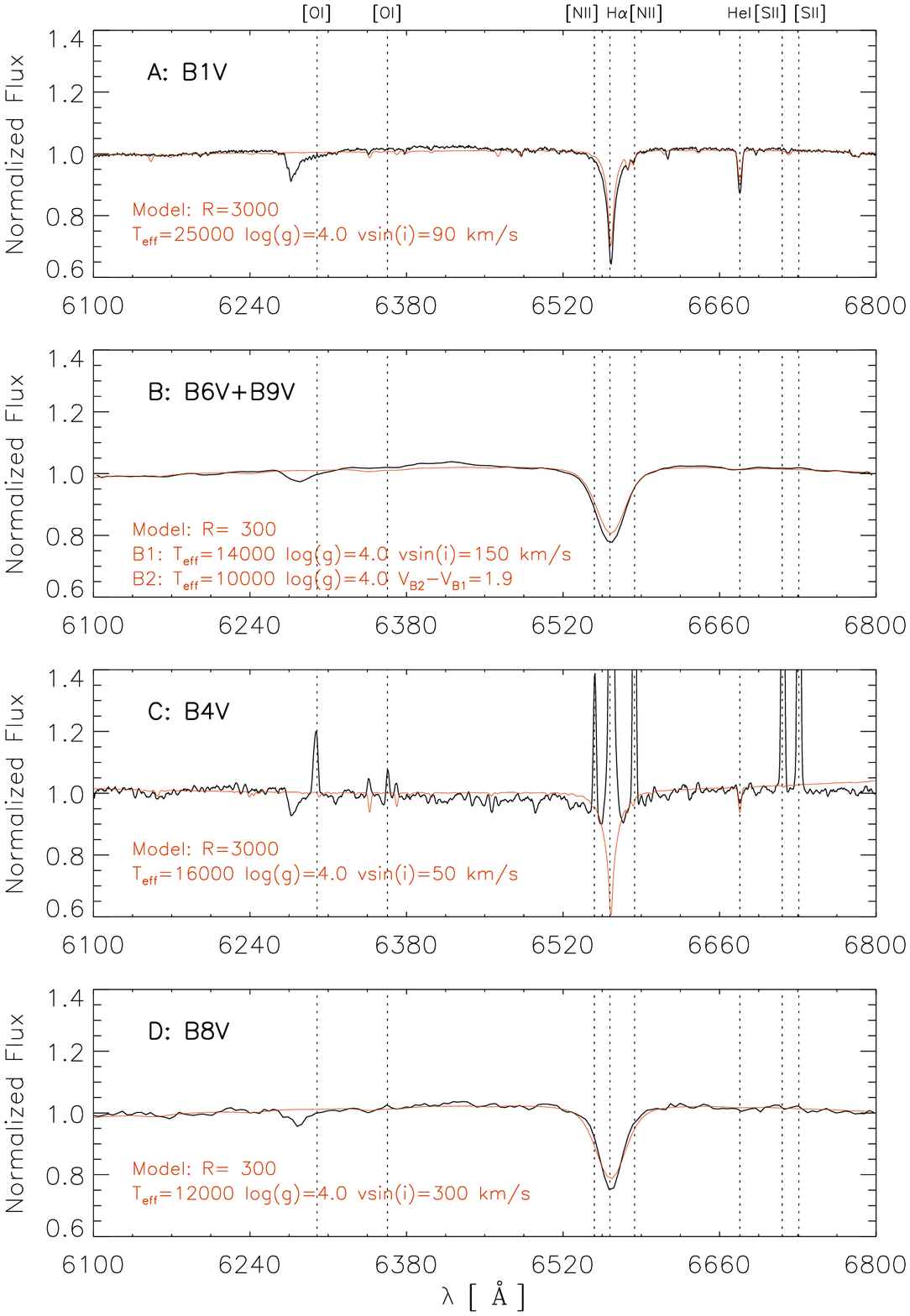}
  \caption{EFOSC2 spectra of the HD~155448 components (black) and 
           rotationally broadened BLUERED 
           synthetic spectra (orange) with $T_{\rm eff}$ and $\log(g)$ of 
           the spectral type found. The spectral resolution of the observed 
           spectrum and its model is indicated in each panel. The absorption 
           features observed from 6250 to 6350\,\AA\ are telluric absorption 
           lines.}
  \label{fig:spec-class}
  \smallskip
\end{figure*}

In Table~\ref{table:spec-class} we present the derived spectral types. We 
find that all objects have spectral type B. In Fig.\,\ref{fig:spec-class}, 
we show the EFOSC2 spectra, together with BLUERED synthetic spectra with 
$T_{\rm eff}$ and $\log(g)$ of the derived spectral type. In the following 
paragraphs, we describe in detail the spectral classification of each source.

\begin{table*}
  \centering
  \caption{Visual to near-IR photometry of the five components of the 
           HD~155448 system.}
  \setlength\tabcolsep{17.9pt}
  \begin{tabular}{lcccccr}
    \hline
    \hline
    \noalign{\smallskip}
    Filter                   &  A         
                             &  B1        
                             &  B2      
                             &  C   
                             &  D 
                             &  Instrument \\
                             &  [mag]
                             &  [mag]
                             &  [mag]
                             &  [mag]
                             &  [mag]
                             &  \\
    \noalign{\smallskip}
    \hline
    \noalign{\smallskip}
    $B$                      &  9.1  $\pm$ 0.1 
                             & \multicolumn{2}{c}{13.0 $\pm$ 0.1}
                             & 17.5  $\pm$ 0.1  
                             & 11.80 $\pm$ 0.05 
                             & EFOSC2 \smallskip \\
    $V$                      &  (1)  
                             & \multicolumn{2}{c}{12.4 $\pm$ 0.1}
                             & 16.1  $\pm$ 0.1  
                             & 11.60 $\pm$ 0.05 
                             & EFOSC2 \smallskip \\
    $R$                      &  (1)
                             & \multicolumn{2}{c}{12.2 $\pm$ 0.1}
                             & 15.0  $\pm$ 0.1  
                             & 11.45 $\pm$ 0.05 
                             & EFOSC2 \smallskip \\
    \hline
    \noalign{\medskip}
    $J$                      &  8.7 $\pm$ 0.1 
                             & 11.6 $\pm$ 0.1  & (2)
                             & 11.9 $\pm$ 0.1  
                             & 11.1 $\pm$ 0.1 
                             & SOFI \smallskip \\
    $H$                      &  8.8 $\pm$ 0.1 
                             & 11.7 $\pm$ 0.1  & (2)
                             & 10.8 $\pm$ 0.1  
                             & 11.1 $\pm$ 0.1 
                             & SOFI \smallskip \\
    NB\_1.64                 &  8.5 $\pm$ 0.1 
                             & 11.6 $\pm$ 0.1  & 13.4 $\pm$ 0.1 
                             & 10.7 $\pm$ 0.1  
                             & 11.0 $\pm$ 0.1 
                             & NACO \smallskip \\
    $K$                      &  8.7 $\pm$ 0.1 
                             & 11.7 $\pm$ 0.1  & (2)
                             &  9.7 $\pm$ 0.1  
                             & 11.1 $\pm$ 0.1 
                             & SOFI \smallskip \\
    NB\_2.12                 &  8.5 $\pm$ 0.1 
                             & 11.5 $\pm$ 0.1  & 13.1 $\pm$ 0.1 
                             &  9.4 $\pm$ 0.1  
                             & 10.9 $\pm$ 0.1 
                             & NACO \smallskip \\
    \noalign{\smallskip}
    \hline
  \end{tabular}
  \\[2mm]
  \flushleft
  Components 
  B1 and B2 cannot be resolved in the visual. Footnotes: (1)~saturated, 
  (2)~not possible to centre a photometric aperture on B2 due to spatial
  resolution. Please see Appendix~\ref{sect:observation} 
  for details about the photometry and its restrictions.
  \label{table:single-flux-1}
  \medskip
\end{table*}

\paragraph{{\bf HD 155448 A:}}

The presence of \ion{He}{i} in absorption lines and the lack of \ion{He}{ii} 
indicates that HD~155448~A has a spectral type B. The lack of the 
\ion{He}{ii} line at 4686\,\AA\ indicates that the star has a spectral type 
later than B0. Since the \ion{Mg}{ii} line at 4481\,\AA\ is much weaker than 
the \ion{He}{i} line at 4471\,\AA, the spectral type of HD 155448 A  should 
be earlier than B5. The presence of the \ion{Si}{iii} lines at 4553, 4568, 
and 4575\,\AA, the \ion{N}{ii} line at 4631\,\AA, the \ion{O}{ii} lines at 
4639 and 4642\,\AA, and the  \ion{C}{iii} lines at 4650\,\AA\ indicate 
that HD~155448~A has a spectral type earlier than B1 if it has luminosity 
class V, and spectral type earlier than B2 if it is a giant. The strength of 
the \ion{O}{ii}, \ion{N}{ii}, and \ion{C}{iii} lines is too weak, and the 
width of the \ion{He}{i} lines and hydrogen lines (H${\beta}$) is too broad 
to be consistent with luminosity classes I and II. While the strength of
\ion{O}{ii}, \ion{N}{ii}, and \ion{C}{iii} lines is consistent with luminosity 
classes III and V, the width of the \ion{He}{i} lines and H${\beta}$ is 
broader than observed in stars of luminosity class III of spectral types B1 
to B2. Therefore, HD~155448~A has a luminosity class V. Altogether,
HD~155448~A has a spectral type B1V. We checked several values of 
$\upsilon \sin(i)$, while the model spectrum giving the best fit suggested 
a $\upsilon \sin(i)$\,=\,90$\pm$5~km/s. The error represents the range of 
$\upsilon \sin(i)$ values that provide a comparable fit to the line profile 
shapes. The mean value given is the one displaying the smallest residuals, 
but there are a few values that are consistent with the observed spectrum.

\paragraph{{\bf HD 155448 B (B1+B2):}}

Because the components B1 and B2 are spatially unresolved, we consider them first 
as a single source. Given that component B1 is at least one magnitude brighter
than component B2, the spectrum of HD~155448~B is dominated by B1. Since 
\ion{He}{i} lines are present at 4471 and 4923\,\AA, and no \ion{He}{ii} 
lines are observed, HD~155448~B should have a spectral type B. Very few 
\ion{He}{i} lines in the range 4450--5100\,\AA\ are present. Therefore, 
HD~155448~B should be a late B-type star. The \ion{He}{i} line at 4471\,\AA\ 
is just slightly stronger than the \ion{Mg}{ii} line at 4481\,\AA. This 
indicates that HD~155448~B should have an earlier spectral type than B7 but 
later than B4. The (weak) strength of the \ion{He}{i} lines at 4923\,\AA\ and 
the barely visible \ion{He}{i} line at 4713 and 5015\,\AA\ indicate that 
HD~155448~B has spectral type B6 or later. Therefore, HD~155448~B most likely 
has the spectral type B6. The broad wings of the H${\beta}$ line are not 
consistent with the luminosity classes I, II, III, and IV, even when 
considering 
a $\upsilon \sin(i)$ as high as 300 km/s. In consequence, HD~155448~B 
is a dwarf star. We conclude that HD~155448~B is a star of spectral type B6V.
The $R$$\sim$3000 model spectrum with the best fit suggests a 
$\upsilon \sin(i)$\,=\,150$\pm$50~km/s.

\begin{table}
  \centering
  \caption{Mid-IR photometry of HD~155448~C.}
  \setlength\tabcolsep{6.1pt}
  \begin{tabular}{lccr}
    \hline
    \hline
    \noalign{\smallskip}
    Filter                   &  C    
                             &  C + Arc
                             &  Instrument \\
                             &  [Jy]
                             &  [Jy]
                             &  \\
    \noalign{\smallskip}
    \hline
    \noalign{\smallskip}
    PAH2 (11.25 $\mu$m)      & 0.51 $\pm$ 10\% & 0.92 $\pm$ 10\%
                             & VISIR \smallskip \\
    PAH2\_ref (11.88 $\mu$m) & 0.44 $\pm$ 10\% & 0.55 $\pm$ 10\%
                             & VISIR \smallskip \\
    $Q2$ (18.72 $\mu$m)      & 0.73 $\pm$ 10\% & 1.46 $\pm$ 10\%
                             & VISIR \smallskip \\
    \noalign{\smallskip}
    \hline
  \end{tabular}
  \\[2mm]
  \flushleft
  In the mid-IR narrow-band filters, only the component C is seen. Please 
  see Appendix~\ref{sect:visir-imag} 
  for details about the photometry and its restrictions.
  \label{table:single-flux-2}
\end{table}

The light of HD~155448~B is dominated by B1. The near-IR images show that 
the B2 component is 1--2~magnitudes fainter than B1. In the EFOSC2 spectrum, 
we notice that the wings of the H$\beta$ line and the \ion{He}{I} line at 
4471\,\AA\ are slightly broader than expected for a B6V star even when
including a large $\upsilon \sin(i)$ ($\sim$200 km/s). These spectral 
characteristics allow us to set constraints on the spectral type of the B2 
component. Assuming the same distance towards B1 and B2 -- and the same 
extinction -- the difference in brightness of 1--2~magnitudes indicates that 
B2 can have a spectral type ranging from B8 to A5. Employing the BLUERED 
templates degraded at the EFOSC2 resolution, we constructed synthetic binary 
star spectra with a B6V primary and a B8V to A5V secondary, taking care of 
scaling the fluxes of each component conforming to their absolute magnitude.
Using our software for spectral comparison, we visually compared the 
synthetic binary spectra and our EFOSC2 spectrum of HD~155448~B. We found 
that the spectrum B6V+B8V is ruled out because the resulting \ion{Mg}{ii} 
line at 4481\,\AA\  is stronger than the \ion{He}{i} line at 4471\,\AA. The 
spectra B6V+A1V to B6V+A5V were ruled out because the secondary is too faint 
to have a noticeable effect on the wings of the H${\beta}$ line. We found 
that the combinations B6V+B9V and B6V+A0V offered the best fits to the 
H${\beta}$ line and the relative strengths of the \ion{Mg}{ii} and 
\ion{He}{i} lines. As the smallest residuals were displayed by the B6V+B9V 
binary spectrum, we suggest that the spectral type of the B2 component is 
B9V. Several values of $\upsilon \sin(i)$ between 100 and 200~km/s gave 
similar solutions, so we assumed for component B2 a $\upsilon \sin(i)$ of 
150$\pm$50~km/s as derived for component B1.

\paragraph{{\bf HD 155448 C}}
exhibits \ion{He}{i} lines in absorption but no \ion{He}{ii} lines. 
Therefore, HD~155448~C has a spectral type B. As the \ion{He}{i} lines have
an FWHM of $\sim$240 km/s, HD~155488~C is likely rotating fast, at least at 
$\upsilon \sin(i)$ of 150 km/s. Emission lines are observed most 
notably H$\alpha$, a narrow emission component inside the H$\beta$ line,
and [\ion{O}{i}], [\ion{N}{ii}], and [\ion{S}{ii}] forbidden emission (see 
Table~\ref{table:emission-lines}). The lack of a strong \ion{He}{i} line in 
absorption at 4713\,\AA\ indicates that HD~155448~C should be later than B1. 
At our $S/N$ there is no evidence of strong \ion{Si}{iii} lines at 4553, 
4568, and 4575\,\AA, and \ion{C}{iii} lines at 4647 and 4651\,\AA. 
Therefore, HD~155448~C should have a spectral type later than B2. As the 
\ion{Mg}{ii} at 4481\,\AA\ line is weaker than the \ion{He}{i} line at 
4471\,\AA, HD~155448~C has a spectral type earlier than B7. In addition, 
the relative strength between these two lines indicates a spectral type 
later than B3. The strength of the \ion{He}{i}, \ion{Mg}{ii}, and the 
H${\beta}$ line are best matched with the spectral type B4. The lack of 
\ion{Si}{iii} lines at 4553 and 4568\,\AA\ and the broad H$\beta$ line 
wings rule out the luminosity classes I, II, and III. The lack of a strong 
\ion{He}{i} line at 5047\,\AA\ is not consistent with luminosity classes 
III and IV. Considering all these characteristics, we conclude that 
HD~155448~C has a spectral type B4Ve. The model exhibiting the best 
fit suggests a $\upsilon \sin(i)$\,=\,50$\pm$10~km/s.

\paragraph{{\bf HD 155448 D:}}

The spectra of HD~155448~D display broad \ion{He}{i} lines 
($\upsilon \sin(i)$\,$\sim$\,370~km/s). Given that no \ion{He}{ii} absorption 
lines are present, HD~155448~D has a spectral type B. The \ion{He}{i} line 
at 4471\,\AA\ and the \ion{Mg}{ii} line at 4481\,\AA\ have a very similar 
strength (with \ion{Mg}{ii} just slightly stronger). This indicates that  
HD~155448~D has a spectral type later than B6 but earlier than B8. The 
strength of the \ion{He}{i} lines at 4923 and 5015\,\AA\ is too weak to be 
consistent with the spectral types B6 and B7, thus HD~155448~D should have 
a spectral type B8. Taking the rotational broadening into account, the width 
of the wings of the H${\beta}$ line is too large to be consistent with  
luminosity classes I, II, III, and IV. As the luminosity class V is the one 
that matches the observed spectrum best, we conclude on a spectral type B8V 
for HD~155448~D. The $R$$\sim$3000 model with the best fit suggests a
$\upsilon \sin(i)$\,=\,300$\pm$15~km/s.

\begin{table}[t]
  \centering
  \caption{Derived spectral types of the HD~155448 components, their
           corresponding absolute magnitudes, colours, and estimated
           distance ranges. See Sect.\,\ref{distances} for details on the 
           determination of $A_V$ and $d$.}
  \setlength\tabcolsep{6.8pt}
  \begin{tabular}{cccccc}
    \hline
    \hline
    \noalign{\smallskip}
    Star   &  Type    &  $M_V$         &  $(B - V)_0$    &  $A_V$      &
                         $d$ (kpc) \\
    \noalign{\smallskip}
    \hline
    \noalign{\smallskip}
    A$^a$  &  B1V     &  $-$3.2$\pm0.6$  & $-$0.27$\pm0.02$  & 1.0$\pm$0.5 &
                         1.7\,$^{+0.7}_{-0.5}$ \smallskip \\
    B1$^b$ &  B6V     &  $-$0.9$\pm0.3$  & $-$0.15$\pm0.02$  & 2.3$\pm$0.9 &
                         1.9\,$^{+1.0}_{-0.7}$ \smallskip \\
    B2$^c$ &  B9V     &  $+$0.2$\pm0.5$  & $-$0.08$\pm0.05$  & 2.3$\pm$0.9 &
                         1.9\,$^{+1.2}_{-0.7}$ \smallskip \\
    C      &  B4Ve    &  $-$1.4$\pm0.2$  & $-$0.19$\pm0.02$  & 5.2$\pm$0.5 &
                         2.8\,$^{+0.7}_{-0.6}$ \smallskip \\
    D      &  B8V     & $-$0.3$\pm0.4$   & $-$0.11$\pm0.02$  & 0.6$\pm$0.5 &
                         1.8\,$^{+0.6}_{-0.4}$ \smallskip \\
    \noalign{\smallskip}
    \hline
  \end{tabular}
  \\[2mm]
  \flushleft
  Notes: The errors given for $M_V$ and $(B-V)_0$ correspond to an error
  in the spectral type of one subclass.\\
  $^a$ Given that the $V$ band image of HD~155448~A is saturated, to estimate
  $V$, we proceeded as described in Sect.\,\ref{distances}.\\
  $^{b,~c}$ Since the components B1 and B2 are unresolved in the $B$ and
  $V$ images, we derived $B$ and $V$ from the observed combined magnitude 
  as described in Sect.\,\ref{distances}.
  \label{table:spec-class}
  \medskip
\end{table}

\subsection{Emission lines in HD~155448~C}
\label{sect:emission-lines}

Emission lines are only seen in the C component. Their intensity, in general,
is stronger for the slit orientation covering the full extent of the arc, 
slit B-C (see Fig.\,\ref{fig:slits}), indicating that the circumstellar 
matter to the northeast is also a source of the emission lines. H${\alpha}$ 
(single-peaked) represents the strongest line, while the forbidden lines 
[\ion{S}{ii}], [\ion{N}{ii}], and [\ion{O}{i}] are fainter. In the following
analysis of the emission lines and their spatial extension, we focus on the 
highest resolution EFOSC2 data, observed in August 2008 with Grism~20
in an east-west slit orientation, because a contamination by the brighter 
component cannot be excluded for the A-C slit orientation.

As $v_\mathrm{rad}$ is unknown, the spectrum is centred such that \ion{He}{i} 
at 6678\,\AA\ -- the only photospheric line in Grism~20 -- lies at v~=~0~km/s.
This results in a velocity shift of ($v_\mathrm{rad} + v_{\sun}$) = +28~km/s. 
The error in the fit of a Gaussian to the \ion{He}{i} line was 9~km/s. Together 
with the general wavelength calibration uncertainty of $\pm$4~km/s, the 
absolute error in the line position is then
$\sqrt{4^2+ 9^2} \sim \pm$10~km/s, while the relative error between two lines
is $\sqrt{4^2 + \chi^2}$, with a typical error $\chi$ in the line centre of 
$<$2~km/s, resulting in a relative error of $\sim$4~km/s. 

To analyse the spatial extension of the emission lines in the eastern direction, we 
calculated position-velocity diagrams, from which we subtracted the continuum 
PSF (see Fig.\,\ref{fig:pos-vel}). The derived line velocities, widths, and 
their spatial offset and extension in the eastern direction are given in 
Table~\ref{table:emission-lines}. We summarise our analysis of the HD~155448~C 
emission lines as follows.

\begin{figure}[t]
  \centering
  \includegraphics[scale=0.34, angle=0]{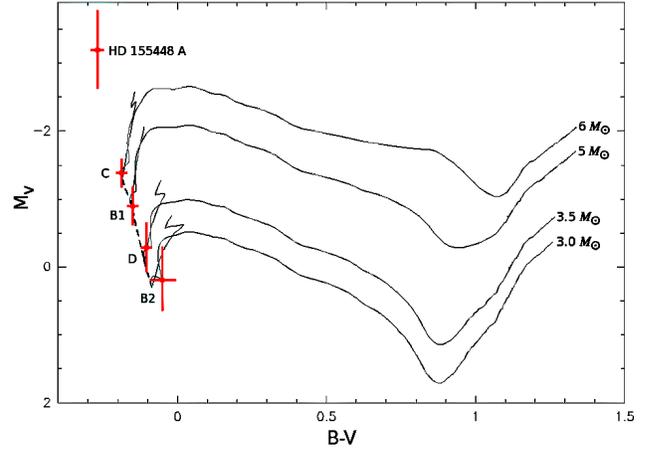}
  \caption{PMS and main sequence evolutionary tracks according to Siess et 
           al.\ (\cite{Siess}) for initial stellar masses corresponding to 
           the spectral types of HD~155448~B1, B2, C, and D. Overplotted are 
           $M_V$ and colours of all components (red crosses, 
           including error bars). The dashed line represents the zero age main 
           sequence. See Sect.\,\ref{age} for further explanations.}
  \label{fig:tracks}
  \bigskip
  \smallskip
\end{figure}

\begin{itemize}

\item{
All emission lines are spatially extended in the eastern direction up to 
$\sim$4\arcsec. The [\ion{O}{i}] line at 6300\,\AA\ is significantly less 
extended than the other lines, while the [\ion{O}{i}] line at 
6363\,\AA\ is unresolved.
}

\item{
All emission lines appear to have a similar velocity shift of $\sim$10~km/s, 
except [\ion{O}{i}] at 6300\,\AA, which is around $\sim$0~km/s. Given that 
the error on the relative position of the emission lines is less than the 
observed shift (see details above), this shift in the forbidden emission lines 
is most likely real. Within their errors, [\ion{S}{ii}], [\ion{N}{ii}], and 
H$\alpha$ are at the same velocity. 
}

\item{
The [\ion{O}{i}] line at 6363\,\AA\ shows a large error in its velocity, 
caused by the weakness of this line. We do not discuss this line in the 
following, and often refer to the [\ion{O}{i}] 6300\,\AA\ line as [\ion{O}{i}]
instead.
}

\end{itemize}

The analysis 
of the line velocities and spatial offsets suggests two components in the 
circumstellar material. One extended component, traced by the [\ion{N}{ii}], 
[\ion{S}{ii}], and H$\alpha$ lines, may be receding from us, possibly in the 
form of a wind or outflow towards the arc, but a full interpretation of 
the system geometry is difficult (see the end of Sect.\,\ref{sect:arc}). A 
second, less extended component, 
traced by the [\ion{O}{i}] line, is close to 0~km/s, but a motion in our 
direction may also be possible within the error
(cf.\ Table~\ref{table:emission-lines}). As the projected velocities are much 
lower than 100 km/s, which would be a characteristic value for a massive 
outflow or jet, the gas may be moving in a direction close to the plane of 
the sky, but the low velocities could also be intrinsic. We return to a 
discussion of the forbidden emission lines in 
Sect.\,\ref{sect:forbidden-lines}.

\begin{table*}[t]
\centering
\caption{Emission lines observed in HD~155448~C for the east-west slit 
         orientation.}
\setlength\tabcolsep{20.7pt}
\begin{tabular}{clcrrcrr}
    \hline
    \hline
    \noalign{\smallskip}
    $\lambda$  &  Line          & EW     &  FWHM          &  Centre  
                                & Offset &  Extension in E   \\
    (\AA)      &                & (\AA)  &  [km/s]        &  [km/s] 
                                & [$"$]  &  [$"$]            \\ 
    \noalign{\smallskip}
    \hline
    \noalign{\smallskip}
    6300       &  [\ion{O}{i}]  & $-$1.0   &  100 $\pm$  6  &  $-$4 $\pm$ 10
                                & $-$0.24 $\pm$ 0.06        &  $-$2.0       \\
    6363       &  [\ion{O}{i}]  & $-$0.3   &  100 $\pm$ 19  &  +13 $\pm$ 14 
                                & $-$0.24 $\pm$ 0.06        &  unresolved   \\
    6548       &  [\ion{N}{ii}] & $-$1.7   &   91 $\pm$  6  &  +13 $\pm$ 10
                                & $-$0.60 $\pm$ 0.06        &  $-$3.0       \\
    6563       &  H${\alpha}$   & $-$23    &   95 $\pm$  4  &  +13 $\pm$ 10
                                & $-$0.60 $\pm$ 0.06        &  $-$4.5       \\
    6583       &  [\ion{N}{ii}] & $-$4.3   &   91 $\pm$  4  &  +11 $\pm$ 10 
                                & $-$0.84 $\pm$ 0.06        &  $-$3.5       \\
    6716       &  [\ion{S}{ii}] & $-$4.0   &   88 $\pm$  4  &   +9 $\pm$ 10 
                                & $-$0.36 $\pm$ 0.06        &  $-$3.5       \\
    6731       &  [\ion{S}{ii}] & $-$6.1   &   88 $\pm$  4  &  +10 $\pm$ 10 
                                & $-$0.36 $\pm$ 0.06        &  $-$3.5       \\
   \noalign{\smallskip}
   \hline
\end{tabular}
\\[2mm]
\flushleft
The error of the FWHM is given by the {\it relative} wavelength calibration 
uncertainty and the uncertainty of the Gaussian fit. For the line centres, 
the error includes the {\it absolute} wavelength calibration uncertainty (see 
Sect.\,\ref{sect:emission-lines}). The spatial extension was determined via 
position-velocity diagrams (see Sect.\,\ref{sect:emission-lines}). 
\label{table:emission-lines}
\bigskip
\medskip
\end{table*}

Could the arc matter have influenced the spectrum of HD~155448~C? As 
shown in Fig.\,\ref{fig:pos-vel} and in Table~\ref{table:emission-lines}, 
we see that the peaks of most emission lines are inside a region 
corresponding to the PSF of HD~155448~C. At the given spatial resolution, 
this indicates that most of this emission is associated with the star. If 
the forbidden emission was dominated by the arc matter, we would have 
observed a clearly higher spatial offset of the emission line peaks. 
Nevertheless, some contribution of the arc matter can be expected as well. 
In particular, the arc's inner border towards HD~155448~C cannot be defined
clearly, and the forbidden emission lines seem to originate 
in a transition region.

High spatial and spectral resolution spectroscopy is required to determine
the contribution of arc matter and to better constrain the velocities of 
the gas.

\subsection{Age}
\label{age}

In Fig.\,\ref{fig:tracks} we plot the derived $M_V$ and colours from 
Table~\ref{table:spec-class} in a colour-magnitude diagram with PMS and main 
sequence evolution tracks from 
Siess et al.\ (\cite{Siess}). Only the tracks for the initial masses 
corresponding to spectral types of the \mbox{\object{HD 155448}} system are 
shown, which is 3.0~$M_{\sun}$ (B2), 3.5~$M_{\sun}$ (D), 5~$M_{\sun}$ (B1), and 
6~$M_{\sun}$ (C). Because the Siess models are only available up to 
7~$M_{\sun}$, corresponding to spectral type B3, HD~155448~A would be located 
above these tracks, as its mass is higher than 7~$M_{\sun}$. Within the 
errors, all stars are very close to 
the ZAMS, thus showing that the system is not a post-AGB or pre-PN 
object, as stated in the literature.

\subsection{Distance}
\label{distances}

The error on the Hipparcos parallax (1.65$\pm$1.99 mas; Perryman et 
al.\ \cite{Perryman}) is too large to derive a meaningful distance, which 
nominally would correspond to 606\,$^{+.....}_{-331}$~pc, with an open upper 
distance limit. Therefore, to constrain the distance of the HD~155448
components, we have to rely on the derived spectral types and photometry. 
In Table~\ref{table:spec-class}, we list the appropriate absolute magnitudes 
for the corresponding spectral classification, obtained from Schmidt-Kaler 
(\cite{Schmidt}). The estimated reddening $A_V$ was derived from the 
observed $B-V$ colours, the intrinsic colours given by Schmidt-Kaler 
(\cite{Schmidt}), and it assumes a standard reddening law ($R=3.14$) 

\begin{math}
A_v = 3.14 \ [(B - V) - (B - V)_0]\,.
\end{math}

The distance follows from the classical distance modulus:

\begin{math}
m_V - M_V = 5 \ \log(d) -  5 + A_v\,.
\end{math}

Since the $V$ band image is saturated for component HD~155448~A, to estimate
$V$ for component A we used $B-V=0.06$ as given in SIMBAD, and $B=9.05$
derived from our EFOSC2 data. Using the Hipparcos $B-V=0.186$ would result
in a $\sim$0.3~kpc closer distance for HD~155448~A. The components B1 and B2
are not resolved in the EFOSC2 images, but only separated in the infrared.
We estimate their individual $B$ and $V$ magnitudes from the observed combined
magnitude, by using the difference in the expected absolute magnitudes between
spectral types B6V and B9V ($\Delta M_{V} = 1.10$, $\Delta M_{B} = 1.18$). For
example, to derive the $V$ magnitude of each component we used

$V_{\rm{B1}}=V_{\rm{B1+B2}}+2.5~{\rm log}\,[1+10\,^{(\Delta M_{V}/-2.5)}]$

$V_{\rm{B2}}=V_{\rm{B1+B2}}+2.5~{\rm log}\,[1+10\,^{(\Delta M_{V}/2.5)}]$\,.

In this way, we obtained $V=12.7\pm0.2$~mag and $B=13.3\pm0.2$~mag for the 
component B1, and for component B2, $V=13.8\pm0.2$~mag and 
$B=14.5\pm0.2$~mag.

We summarise the results of the distance determination in 
Table~\ref{table:spec-class}. The distances to all components are consistent 
within their errors, with a mean distance of 2.2~kpc. While the uncertainty 
in the distance estimate seems large, we emphasise that these error ranges 
are mainly caused by an uncertainty of only one subclass in the spectral type 
derivation.

\section{Results derived from infrared data}
\label{sect:IRanalysis}

\subsection{A fifth component in the \mbox{HD 155448} system}

\begin{figure*}
\centering
\begin{tabular}{cccc}
  \includegraphics[scale=.407, angle=0]{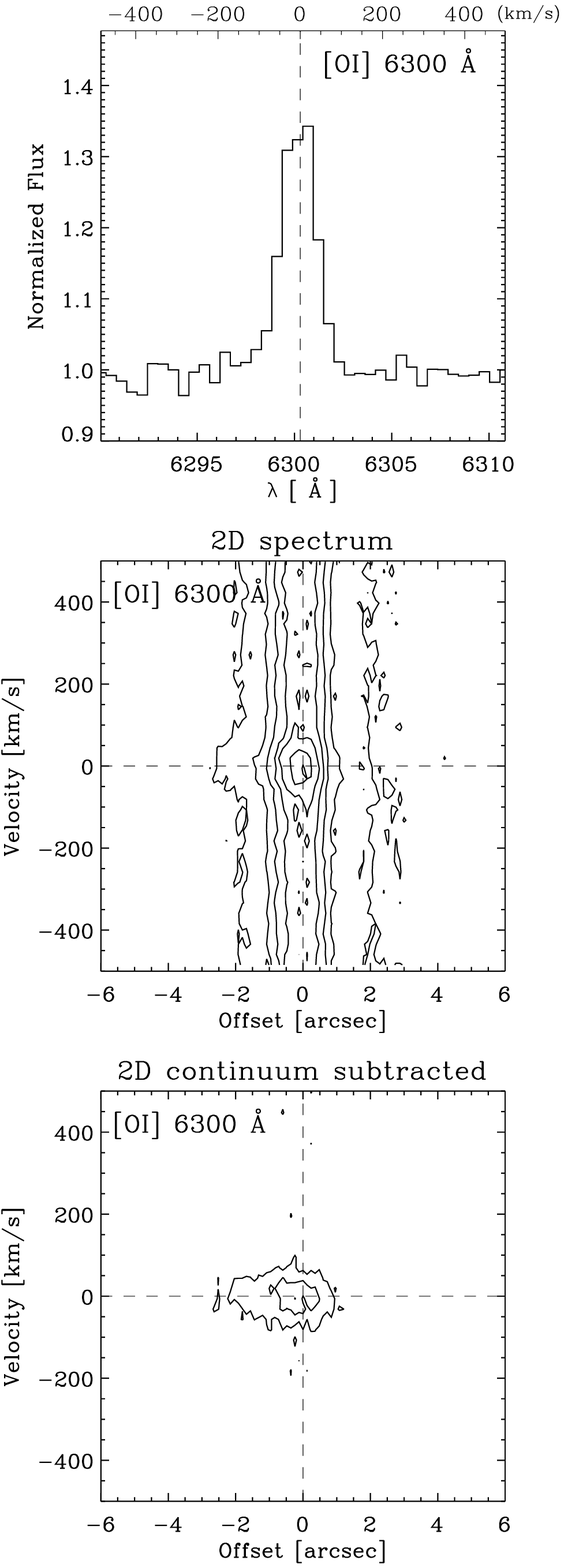} &
  \includegraphics[scale=.407, angle=0]{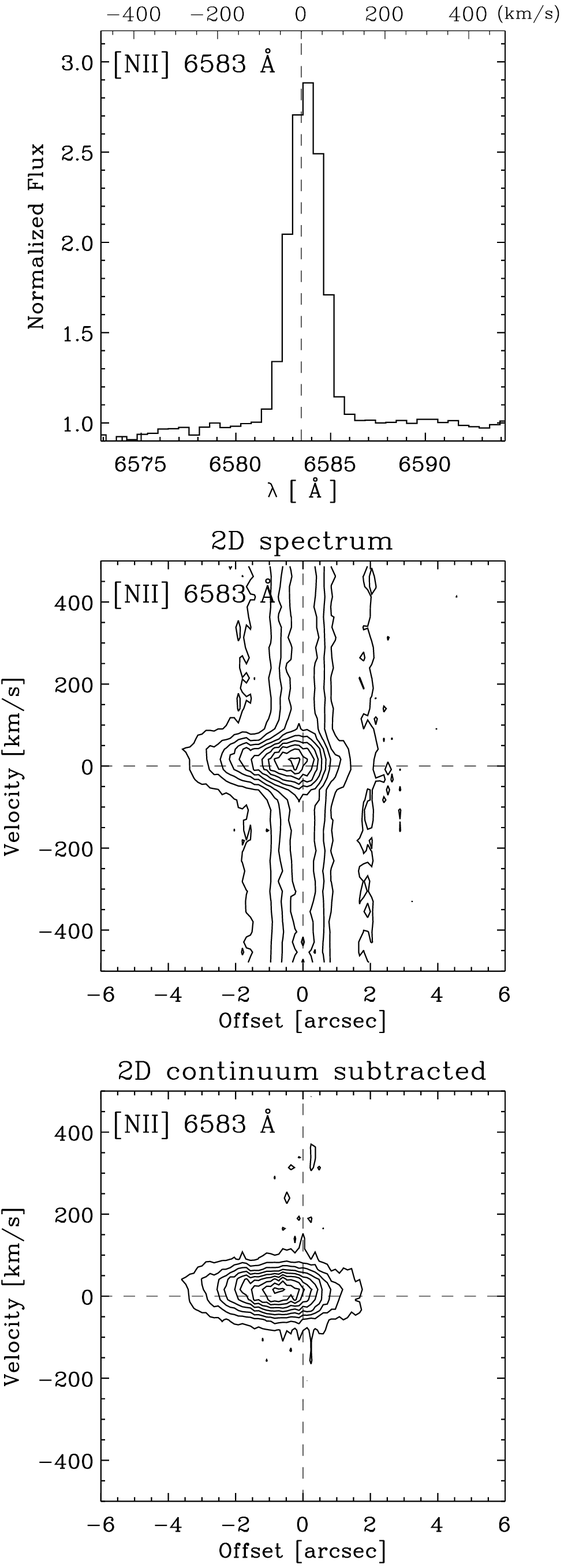} &
  \includegraphics[scale=.407, angle=0]{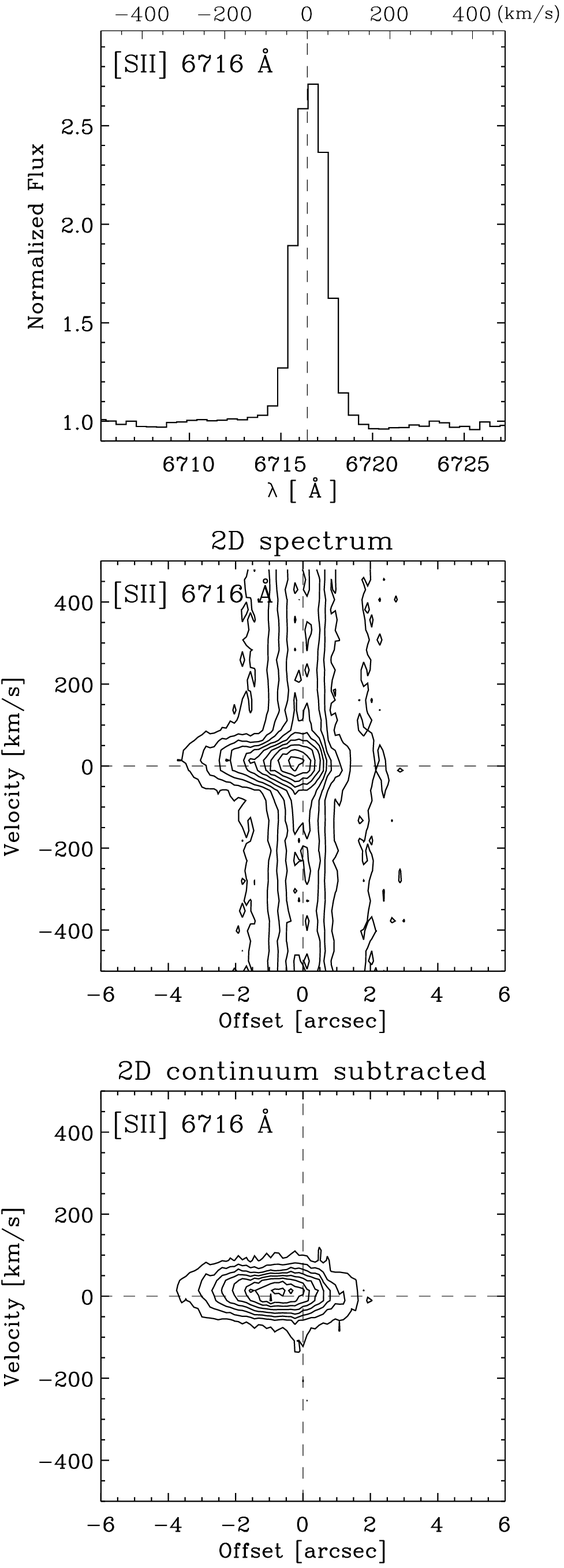} &
  \includegraphics[scale=.407, angle=0]{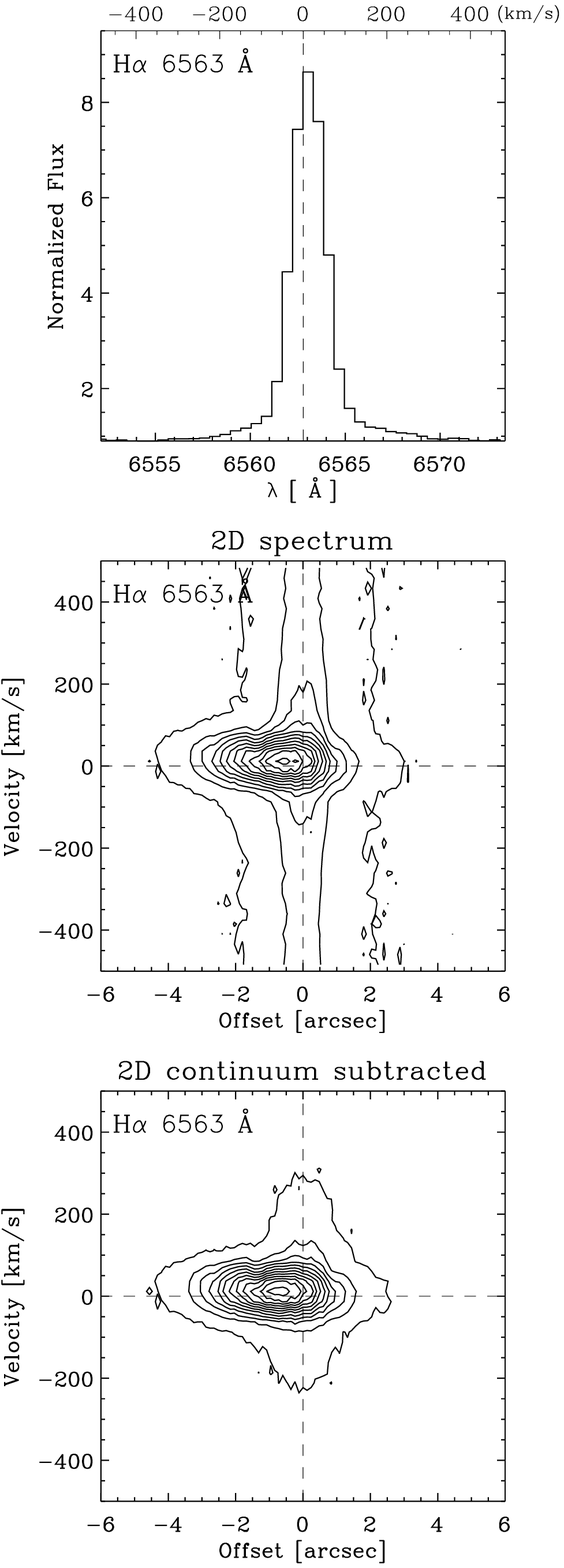} \\
\end{tabular}
\caption{Important emission lines observed in HD 155448 C, extracted from
         spectra obtained in an east-west slit orientation.
         The upper panel displays the extracted spectral line, the 
         middle panel shows the position-velocity diagrams in the raw 
         2D spectrum, while the lower panels expose the 
         position-velocity diagrams after the subtraction of the continuum 
         PSF. The first (outer) contour represents the 5\,${\sigma}$ level 
         with respect to the background noise. Subsequent contours are at 
         10\,${\sigma}$ intervals, except for H$\alpha$, which is composed of 
         30\,${\sigma}$ intervals.}
\label{fig:pos-vel}
\bigskip
\medskip
\end{figure*}

To our knowledge, this paper offers the first high-resolution, 
adaptive optics observations of the \mbox{HD 155448} system. The ADONIS 
data reveal that the component \mbox{HD 155448 B} can be resolved into two 
stars (cf.\ Fig.\,\ref{fig:slits}) with a separation of 1.21$\pm$0.04$''$. 
This additional companion is also visible in NACO and SOFI data. In NACO
data, the separation is 1.16$\pm$0.01$''$, and 1.14$\pm$0.05$''$ for SOFI. 
Subpixel accuracy was applied in calculating the component separations. 
We do not want to speculate about whether we see any motion during the five
years between 
these data sets, because all results (marginally) overlap within their error 
ranges. The photometry is given in Table~\ref{table:single-flux-1}.

\subsection{Imaging and photometry}

In the images we found arc-shaped emission, curving from HD~155448~C 
towards the northeast. This feature is seen both in the near- and mid-IR, and 
is also visible in optical data. The extension is up to $\sim$3\arcsec\ from 
HD~155448~C in all the observed passbands. A difference in length of 
0.2\arcsec\ between the filters may depend on the sensitivity of the 
individual data sets. The inner end of the arc, facing towards HD~155448~C, 
is not clearly constrained. At optical wavelengths, in the near-IR and in the 
mid-IR around 10~$\mu$m, little extra emission is seen very close to the star, 
while significant emission starts at distances beyond $\sim$1.0\arcsec\ from 
HD~155448~C. In the mid-IR PAH band filter and at shorter wavelengths, the 
intensity is maximal in the arc's centre. The appearance in the $Q$ band 
filter is clearly different, where the arc extends all the way towards the 
star. Moreover, the peak of the emission is now located close to the star, 
in an area that shows nearly no emission in the near-IR or at visual 
wavelengths.

Tables~\ref{table:single-flux-1} and~\ref{table:single-flux-2} list the 
photometry for all passbands used
in this work. Only for adaptive optics data was it possible to obtain 
resolved photometry for the B2 companion. The SOFI and NACO magnitudes 
agree well within their errors, despite the problems calibrating the NACO 
data (see Appendix~\ref{naco-imag}). For VISIR we performed PAH-on 
(11.25~$\mu$m) and PAH-off differential
imaging (continuum around 11.88~$\mu$m), and subtracted both frames after a 
re-centring with 0.1~pixel accuracy. For a description of the errors in 
the photometry, we refer to Appendix~\ref{sect:visir-imag}. In 
Table~\ref{table:single-flux-2} we also give for VISIR the total flux of 
HD~155448~C plus the circumstellar matter in the arc, because resolving both 
is not accurately possible in the $Q$ band, as is recognisable in 
Fig.\,\ref{fig:mir_imag}.

\subsection{Mid-IR spectroscopy}

\subsubsection{VISIR}
\label{sect:mir-spec-visir}

\begin{figure*}[t]
  \centering
  \includegraphics[scale=.885, angle=0]{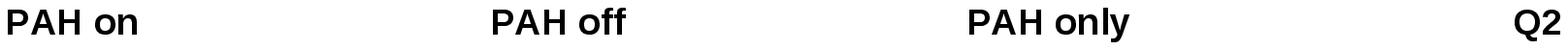} \\ 
  \includegraphics[scale=.855, angle=0]{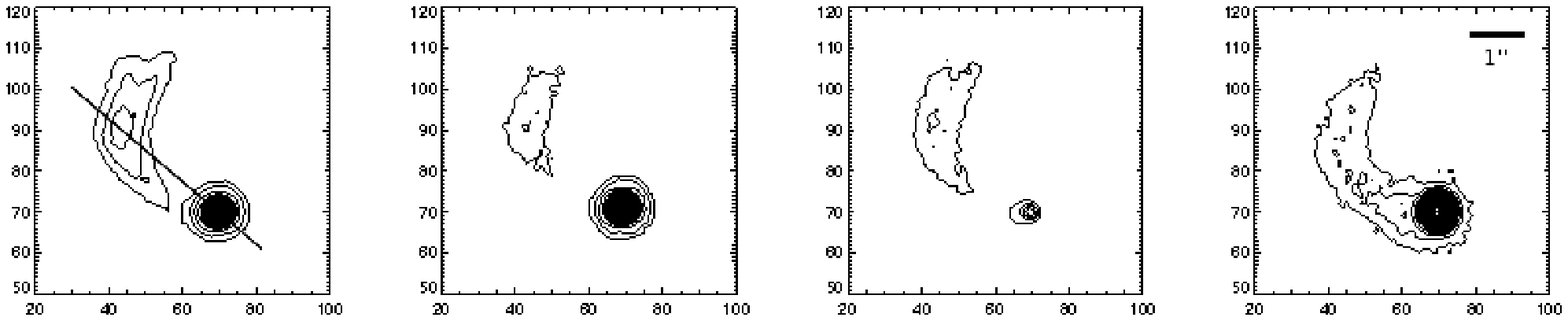}
  \caption{VISIR images in the PAH2 filter at 11.25~$\mu$m (``PAH on''), the 
           continuum at 11.88~$\mu$m (``PAH off''), the difference of both 
           previous images (``PAH only''), and in the $Q2$ filter at 
           18.72~$\mu$m. Only the C component is seen in the mid-IR narrow-band
           filters used. The VISIR slit orientation at 50$^\circ$ position 
           angle is indicated in the left image (cf.\ also 
           Fig.\,\ref{fig:irs-slits}). To obtain optimal contour plots 
           for the different intensity ranges, 50 equidistant contour levels 
           were plotted for ``PAH on'' and ``PAH off'', 5 contour levels for 
           ``PAH only'', and 20 levels for the $Q2$ image.}
  \label{fig:mir_imag}
  \bigskip
  \medskip
\end{figure*}

We used VISIR low-resolution spectra to study the composition of the
circumstellar matter close to HD~155448~C and in the arc. None of the other 
stellar components is detected in the mid-IR, what was also seen 
in our previous test observations using the TIMMI2 mid-IR camera at ESO La 
Silla Observatory, which has a larger field of view than VISIR. However, our 
mid-IR narrow-band filters used 
cover only a small -- although representative -- range of the $N$ band. An 
eventual mid-IR emission originating in the remaining HD~155448 components 
at other $N$ band wavelengths cannot be fully excluded. 
The slit orientation and the procedure for disentangling the emission near the 
star from that of the arc is explained in Appendix~\ref{sect:visir-spec}.

\begin{figure*}[t]
  \centering
  \includegraphics[scale=0.50, angle=0]{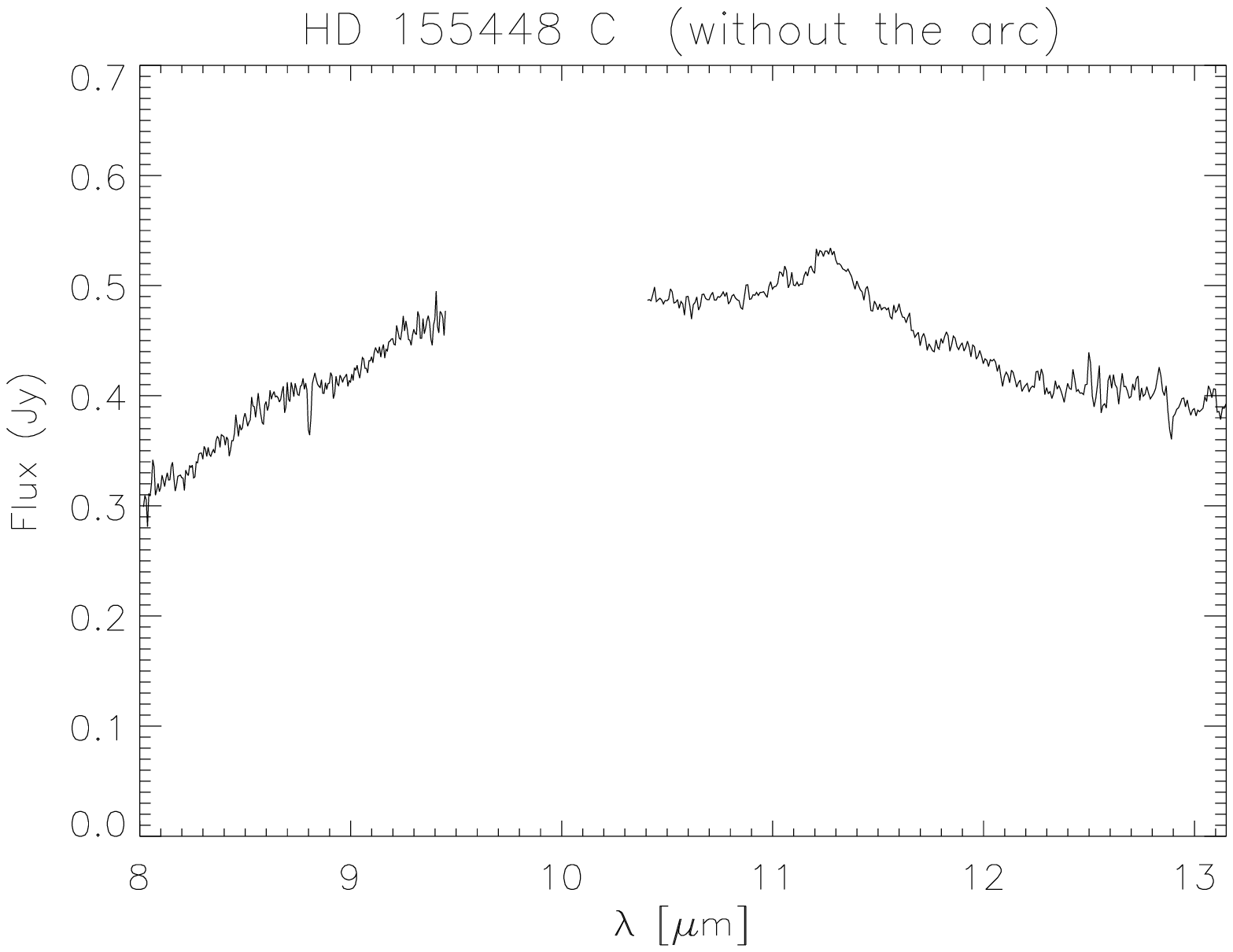}
  \includegraphics[scale=0.50, angle=0]{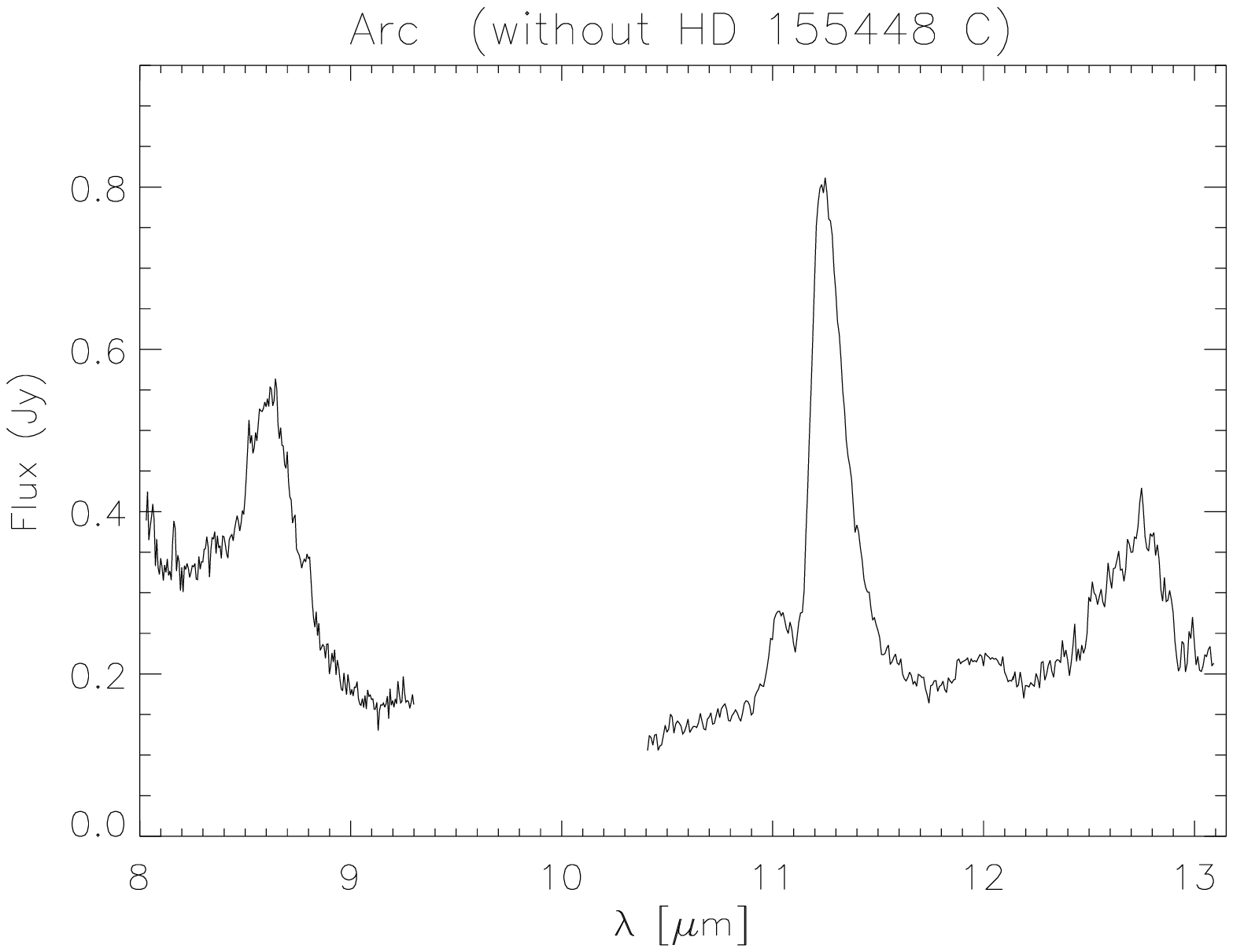}
  \caption{VISIR $N$ band spectra. Left panel: HD~155448~C, showing a
           silicate feature. Right panel: the arc, with strong PAH bands at 
           8.6, 11.3, and 12.7~$\mu$m. The y-axis scale is not 
           identical in both figures, but was adjusted to optimally show 
           the features. The gap between approximately 9.5~$\mu$m and 
           10.3~$\mu$m is due to incomplete instrument coverage (see 
           Appendix~\ref{sect:visir-spec}).
           }
  \label{fig:mir_spec}
  \bigskip
  \medskip
\end{figure*}

\paragraph{{\bf HD~155448~C}}:
The spectrum of HD~155448~C is dominated by silicate emission between
$\sim$8--12~$\mu$m (Fig.\,\ref{fig:mir_spec}, left panel). For a description 
of silicate emission features at these wavelengths, we refer, e.g., to 
Sch\"utz et 
al.\ (\cite{Schuetz}) and references therein. The silicate feature of
HD~155448~C can be interpreted as emission mainly from small, amorphous
silicate grains located in the hot surface layer of a circumstellar disk.
A smaller peak at 11.3~$\mu$m indicates either the presence of crystalline 
silicates or of polycyclic aromatic hydrocarbons (PAHs), which both exhibit an
emission peak around this wavelength. We provide more details on the dust
composition in Sect.\,\ref{sect:mir-spec_spitzer}.

\begin{figure}[t]
  \centering
  \includegraphics[scale=0.51, angle=0]{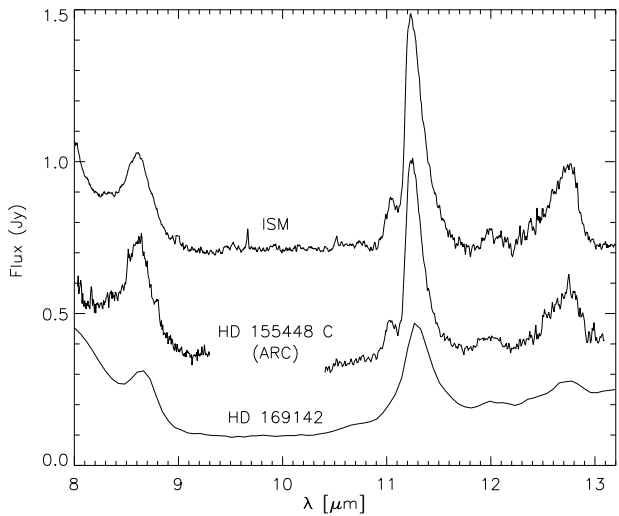}
  \caption{Comparison of the PAH bands in the mean ISM (from Hony et 
           al.\ \cite{Hony}), in \object{HD 169142} (IRS spectrum), and in
           the arc near HD~155448~C. The feature at 11.0~$\mu$m clearly 
           differentiates the arc matter near HD~155448~C from HD~169142, 
           while the feature is also present in the ISM.}
  \label{fig:pah_comp}
  \bigskip
\end{figure}

\paragraph{{\bf Arc}}:
Unlike for the spectrum of the stellar component, we see no silicate features 
related to the arc. Here, the spectrum is dominated by strong PAH bands at 8.6, 
11.3, and 12.7~$\mu$m. We also see fainter PAH features at 11.0 and 12.0~$\mu$m.
In Table~\ref{tab:pah} we give a detailed list of the measured properties of 
the observed features and of the bending modes that can cause them (from Hony et 
al.\ \cite{Hony}). Based on the spectral appearance of PAH bands, Peeters et 
al.\ (\cite{Peeters}) define characteristic PAH classes A\,/\,B\,/\,C, which
represent different stages of PAH processing. By comparing the peak position 
of the PAH bands in our spectrum to these classes, it is clear that the arc's 
PAH features do not belong to the highest processed class~C, as there is a 
strong feature at 8.6~$\mu$m that is absent in class C sources (Peeters et 
al.\ \cite{Peeters}). To decide on class A or B, we need to also consider
the PAH bands between 6--8~$\mu$m, and refer to the Spitzer spectrum in the 
next section. In Fig.\,\ref{fig:pah_comp}, we compare the PAH spectrum 
of the arc with that of the mean ISM and that of \mbox{\object{HD 169142}}, a 
prototype of PAH features observed in  Herbig Ae/Be stars (see Acke 
et al.\ \cite{Acke2010}). It is clear that the arc spectrum is more like
that of the ISM than of the Herbig Ae star, especially when 
focussing on the region near 11~$\mu$m. The small feature at the base of the 
11.3~$\mu$m band towards shorter wavelengths is the 11.0~$\mu$m PAH 
band, which is attributed to the out-of-plane bending of CH in PAH cations. 
Therefore, its emission is an excellent tracer of ionised interstellar PAHs. 
Indeed, in a large sample of isolated Herbig Ae/Be stars, the 11.0~$\mu$m 
feature is absent (Acke et al.\ \cite{Acke2010}). {\it This suggests that the 
arc-shaped, PAH emitting region is not directly related to the disk around 
HD~155448~C}, but rather an ISM remnant of the stellar formation process.

\subsubsection{Spitzer}
\label{sect:mir-spec_spitzer}

\begin{figure*}[t]
  \centering
  \includegraphics[scale=0.32, angle=0]{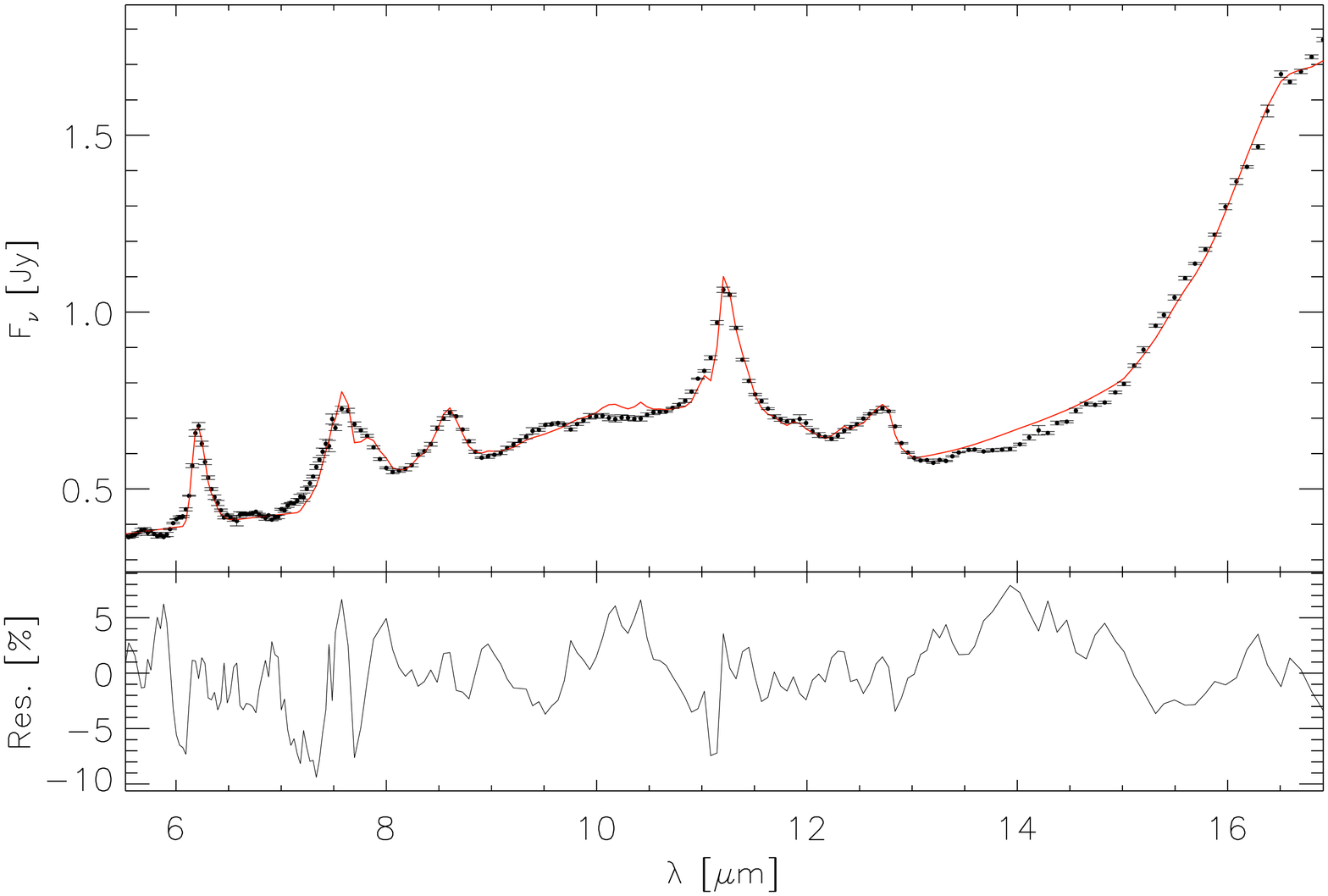}
  \includegraphics[scale=0.32, angle=0]{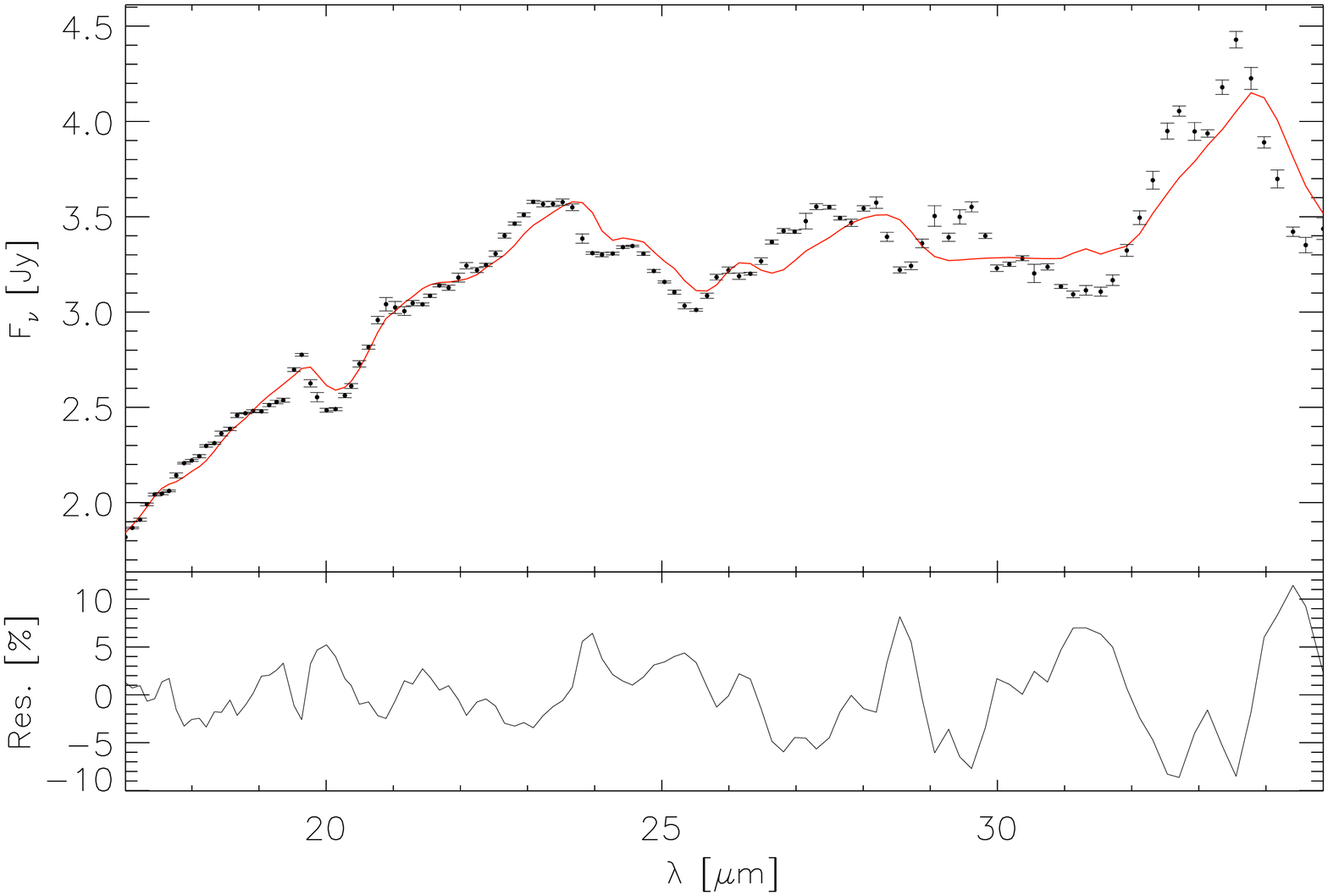}
  \caption{Fit of the Spitzer IRS spectrum of HD~155448~C. The points in the 
           upper panels represent the Spitzer data, while the overplotted 
           continuum line is the fitted model. See Table~\ref{tab:spitzer} 
           for the
           individual contributions of the model dust species. The residuals 
           of the fit are shown in the lower panels. The quality of the fit 
           is good in general; however, there are mismatches between our 
           model and the observed spectra. In the 5--17~$\mu$m region most 
           of the mismatches are related to the strong PAH emission features. 
           For the 17--35~$\mu$m region the mismatches are instead caused by 
           differences in the peak positions of the strong crystalline 
           silicate bands between the model and the observed spectrum.
           }
  \label{fig:spitzer_spec}
  \bigskip
  \medskip
\end{figure*}

As described in Sect.\,\ref{sect:mir-spec-visir}, we see the emission from 
HD~155448~C and the arc in the mid-IR. However, in the
Spitzer spectra the emission from both sources is not spatially resolved.
To derive the composition and physical parameters of the dust 
grains, whose emission features are clearly visible at mid-IR wavelengths, 
we used the two-layer temperature distribution (TLTD) method. For details 
of the spectral analysis method, as well as for the applied optical constants 
of the dust species used, we refer to Juh\'asz et al.\ (\cite{Juhasz}). 
The spectrum was fitted in two wavelength regions: short (5--17~$\mu$m) 
and long (17--35~$\mu$m), with five dust components: amorphous silicates with 
olivine and pyroxene stoichiometry, crystalline forsterite and enstatite, and 
amorphous silica. 
Three different grain sizes were used for the amorphous silicate and silica 
components (0.1, 1.5, and 6.0~$\mu$m), while only two for the crystals (0.1 
and 1.5~$\mu$m), as no evidence was found for large crystals in the 
spectrum. A mean ISM PAH template from Hony et al.\ (\cite{Hony}) was used 
to derive the individual PAH band profiles for the fits. The results are 
shown in Fig.\,\ref{fig:spitzer_spec} and Table~\ref{tab:spitzer}.

\paragraph{{\bf Silicates}}:
The absolute flux level in the Spitzer IRS spectrum is higher in the 
8--13~$\mu$m interval than in the VISIR spectrum, indicating that the 
Spitzer spectra may include more emission from the surrounding 
material due to the wider slit. The dust composition in the 10~$\mu$m 
region is dominated by small (0.1~$\mu$m) amorphous grains. The low 
peak-to-continuum ratio of the 10~$\mu$m silicate feature, which is usually 
explained by the presence of large grains, is probably caused by the high
extinction towards HD~155448~C. Crystalline silicates represent only about 
13\,\% of the total silicate mass in the short interval. This is a normal value 
among Herbig Ae/Be stars (see, e.g., Juh\'asz et al.\ \cite{Juhasz2010} or 
van Boekel et al.\ \cite{Boekel05}). The crystal population is dominated by 
forsterite, whereas the contribution of enstatite is negligible. When comparing 
the fit of the short interval with that of the long interval, we see that the 
forsterite-to-enstatite mass ratio decreases with wavelength (i.e.\ with radius 
in the disk), similar to what is observed in other Herbig Ae/Be systems 
(Juh\'asz et al.\ \cite{Juhasz2010}), and even T~Tauri stars (Meeus et al.\ 
\cite{Meeus}). We note that, although the general quality of the fit 
is good, there are mismatches between the observed and the model spectrum. 
In the short region these mismatches are related to the PAH features, 
because their shape was not perfectly reproduced by our model templates. In 
the long wavelength interval, the crystalline silicate bands in our model 
seem to be systematically shifted towards longer wavelengths than in 
the observed positions. These uncertainties in band positions are likely 
related to grain shape effects and/or to the dependence of the mass-absorption 
coefficients on the temperature that is not included in our analysis (see also 
Juh\'asz et al.\ \cite{Juhasz2010} for a discussion on this topic).

\paragraph{{\bf PAH features}}:
The observed shape and positions of the PAH bands in the spectrum of 
\object{HD 155448 C} are consistent with that of a Class A profile in 
Peeters et al.\ (\cite{Peeters}), typically found in the ISM (cf.\ also 
Sect.\,\ref{sect:mir-spec-visir}, paragraph on the arc).

\begin{itemize}

\item{
The dominance of the 7.6~$\mu$m sub-band over that of the 7.8~$\mu$m 
sub-band in the 7.7~$\mu$m PAH feature complex suggests that the emitting 
molecules are small (number of carbon atoms $N_{\rm C}<$\,90--100) and 
ionised (Bauschlicher et al.\ \cite{Bauschlicher}).
}


\item{
Acke et al.\ (\cite{Acke2010}) find decreasing aliphatic over aromatic 
line strength ratios with increasing stellar effective temperature. This 
correlation is explained by the higher stability of the aromatic bonds 
against the destructive UV photons compared to the aliphatic bonds. Given 
that no evidence was found for aliphatic emission features 
(at 6.8~$\mu$m and 7.2~$\mu$m) in its spectrum, \object{HD 155448} seems 
to follow this trend.
}

\item{
Hony et al.\ (\cite{Hony}) and Keller et al.\ (\cite{Keller}) interpret 
the 6.2/11.2~$\mu$m and 12.7/11.2~$\mu$m line strength ratios as indicators 
of the ionisation state and irregularity of the molecules. Comparing the 
calculated line strength ratios of \object{HD 155448} (F6.2/F11.2 =
1.52\,$\pm$\,0.12, F12.7/F11.2 = 0.42\,$\pm$\,0.04) to Fig.\,5 in Hony et 
al.\ (\cite{Hony}), the PAH molecules in the arc near HD~155448~C are 
probably ionised and irregular, pointing to PAH processing by the relative 
intensive UV field in the HD~155448 environment.
}

\end{itemize}

\section{Position and motion}
\label{comotion}

We do not aim to give absolute astrometry, since the coordinates 
were not derived by means of astrometric standard stars. At the time of 
our observations, both ADONIS and EFOSC2 data had not yet been offered with 
a world coordinate system, which we were able to fix for the EFOSC2
files. Obviously, the NACO adaptive optics data permit a much more 
precise calculation of positions and angles than the SOFI frames, but 
the rather short time in between these two data sets (one year) does
not permit calculating component motions at the distance of the 
HD~155448 system. Therefore, we focus only on the NACO data in 
Table~\ref{table:positions}. The errors in our RA and Dec can be determined 
from the positional uncertainty in the data sets of various nights, which were 
taken with different guide stars. (The astrometric accuracy is limited by the 
accuracy of the guide star coordinates.) We do not quote RA and Dec for 
the literature data, as the VizieR web interface points out that their 
accuracy could be low.

\begin{table}[t]
\centering
\caption{Properties of the PAH features.}
\setlength\tabcolsep{5.7pt}
\begin{tabular}{rlrr}
   \hline
   \hline
   \noalign{\smallskip}
   Feature  &  Bending Mode    &  Peak Position  &  Central Wavelength  \\
   ($\mu$m) &                  &    ($\mu$m)     &       ($\mu$m)       \\
   \noalign{\smallskip}
   \hline
   \noalign{\smallskip}
   6.2      &  CC stretching   &      6.21       &         6.23         \\
   7.6      &  CC stretching   &      7.58       &         7.58         \\
   7.8      &  CC stretching   &      7.82       &         7.85         \\
   8.6      &  CH in plane     &      8.64       &         8.66         \\
   11.0     &  CH out of plane &     11.03       &        11.02         \\
   11.3     &  CH out of plane &     11.25       &        11.26         \\
   12.0     &  CH out of plane &     12.01       &        12.01         \\
   12.7     &  CH out of plane &     12.75       &        12.63         \\
   \noalign{\smallskip}
   \hline
\end{tabular}
\\[2mm]
\flushleft
The central wavelength was determined by calculating the weighted centre of 
the feature.
\label{tab:pah}
\bigskip
\medskip
\end{table}

\begin{table}[t]
\centering
\caption{Dust composition of the Spitzer IRS spectrum, in two wavelength 
         regions.}
\linespread{1.3}{
\selectfont
\setlength\tabcolsep{5.6pt}
\begin{tabular}{lrrr}
   \hline
   \hline
   \noalign{\smallskip}
   Parameter           &  Grain           &  Mass fraction/ & Mass fraction/ \\
                       &  size [$\mu$m]   &  PAH intensity  & PAH intensity  \\
                       &                  &  (5--17$\mu$m)  & (17--35$\mu$m) \\
   \noalign{\smallskip}
   \hline
   \noalign{\smallskip}
   $\chi^2$            &               &  61.8                  & 
                                          91.1                 \\
   Amorph.\ Olivine    &  0.1          &  71.2 $^{+2.2}_{-2.3}$   & 
                                          32.1 $^{+3.7}_{-3.9}$   \\
   Amorph.\ Olivine    &  1.5          &  8.2  $^{+2.4}_{-2.2}$   & 
                                          53.1 $^{+4.3}_{-4.5}$   \\
   Amorph.\ Olivine    &  6.0          &  -                     & 
                                          -                     \\
   Amorph.\ Pyroxene   &  0.1          &  -                     & 
                                          -                     \\
   Amorph.\ Pyroxene   &  1.5          &  -                     & 
                                          -                     \\
   Amorph.\ Pyroxene   &  6.0          &  -                     & 
                                          -                     \\
   Cryst.\ Forsterite  &  0.1          &  2.8  $^{+0.4}_{-0.3}$   & 
                                          5.2  $^{+0.3}_{-0.3}$   \\
   Cryst.\ Forsterite  &  1.5          &  10.3 $^{+0.6}_{-0.5}$   & 
                                          -                     \\
   Cryst.\ Enstatite   &  0.1          &  -                     & 
                                          1.9  $^{+0.2}_{-0.2}$   \\
   Cryst.\ Enstatite   &  1.5          &  -                     & 
                                          -                     \\
   Amorph.\ Silica     &  0.1          &  5.3  $^{+0.3}_{-0.3}$   & 
                                          -                     \\
   Amorph.\ Silica     &  1.5          &  1.0  $^{+0.8}_{-0.6}$   & 
                                          5.3  $^{+0.2}_{-1.7}$   \\ 
   Amorph.\ Silica     &  6.0          &  -                     & 
                                          2.3  $^{+1.2}_{-0.9}$   \\
   \noalign{\smallskip}
   \hline
   \noalign{\smallskip}
   PAH 6.2             &               &  0.285 $^{+0.004}_{-0.004}$  & 
                                          -                        \\
   PAH 7.6             &               &  0.227 $^{+0.005}_{-0.004}$  & 
                                          -                        \\
   PAH 7.8             &               &  0.189 $^{+0.003}_{-0.002}$  & 
                                          -                        \\
   PAH 8.6             &               &  0.226 $^{+0.003}_{-0.003}$  & 
                                          -                        \\
   PAH 11.3            &               &  0.352 $^{+0.006}_{-0.007}$  & 
                                          -                        \\
   PAH 12.7            &               &  0.137 $^{+0.003}_{-0.003}$  & 
                                          -                        \\
   \noalign{\smallskip}
   \hline
\end{tabular}
} 
\\[3mm]
\flushleft
Relative silicate mass fractions and their errors are given in percent. The 
mass fractions marked as non-detections are either fitted as 0\% or have 
$<$\,1$\sigma$ significance. The numbers indicated for PAHs are relative 
intensities compared to the normalised PAH model profile from Hony et 
al.\ (\cite{Hony}).
\label{tab:spitzer}
\bigskip
\end{table}

Can we see evidence for a gravitationally bound motion of the 
HD~155448 components? The proper motion of $\delta$RA\,$=$\,2.77~mas/yr and 
$\delta$Dec\,$=$\,$-$0.85~mas/yr, together with their uncertainties of the same 
order (Perryman et al.\ \cite{Perryman}; the catalogue entry refers 
only to HD~155448~A), could -- during the past $\sim$80 years -- have caused
a change in $\rho$ up to $\sim$0.4\arcsec\ for unbound stars, depending on
how the mean proper motion and its uncertainty would coadd. This value could 
eventually be doubled, when the random motion of two stars would be in the
opposite direction. Such a large motion is not seen between the components 
A, B, and C. On the other hand, the mean proper motion and its uncertainties
could also result in lower values. Without knowledge of the precise 
coordinates from the years 1911 and 1928, respectively, we cannot analyse 
the precise trajectories of the components. However, no 
significant angular motion is seen between now and $\sim$80 years ago.

\section{Discussion}
\label{sect:discussion}

\subsection{SED of the HD~155448 system}
\label{sect:discussion-sed}

Our analysis shows that HD~155448 is a young quintuple system, in contrast
to earlier descriptions in the literature referring to it as a post-AGB star. 
This earlier classification was, however, only based on IRAS colours, which 
does not permit a unique classification (cf.\ Sect.\,\ref{sect:introduction}). 

\begin{figure}[t]
  \centering
  \includegraphics[scale=0.31, angle=0]{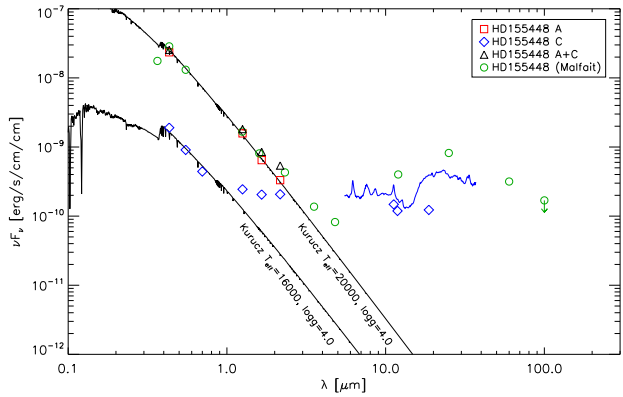}
  \caption{Spectral energy distributions of the HD~155448~A and C components 
           constructed from spatially resolved photometry. Overplotted are the 
           fitted Kurucz atmosphere models. See Sect.\,\ref{sect:discussion-sed}
           for an interpretation.}
  \label{fig:resolved-seds}
  \bigskip
  \medskip
\end{figure}

Based on the interpretation of the SED, Malfait et al.\ (\cite{Malfait}) derived
the presence of a large inner hole in the disk of HD~155448~A. However, due 
to missing spatial resolution in the photometry, their SED merged the fluxes 
of HD~155448~A, dominating the visual wavelengths, and the mid-IR emission
from HD~155448~C. In Fig.\,\ref{fig:resolved-seds}, we plot 
resolved SEDs for both components. Our photometry is dereddened and plotted 
in comparison to a Kurucz atmosphere model (Kurucz \cite{Kurucz}). The sum 
of the spatially resolved photometry of HD~155448~A and C agrees well with 
the spatially unresolved data of Malfait et al.\ (\cite{Malfait}). This means 
that, instead of a single transitional disk, the HD~155448 system hosts four 
stars with very little or no circumstellar material and one with a disk. 
Substantial NIR excess emission is seen for HD~155448~C. If this emission 
indeed originates in a circumstellar disk, there is no reason to assume an 
inner hole, but rather a hot inner rim. 

Besides the quintuple, we uncovered an arc-shaped emission region 
northeast of HD~155448~C.
We exclude the possibility that the arc could be a chance alignment because 
in the $Q$ band, we see that the circumstellar matter is connected with the 
star (Fig.\,\ref{fig:mir_imag}). But what is the origin of this circumstellar 
matter and what causes the observed elongated curved emission?

\subsection{Origin of the forbidden emission lines}
\label{sect:forbidden-lines}

For late-B and A-type Herbig stars, the formation of a circumstellar 
disk during the stellar formation process is theoretically predicted. The 
observed geometry of the circumstellar matter is less clear for early-B stars, 
where the material may be present both in spherical (envelope) and disk-like 
geometries. The classical Be star phenomenon is 
explained by ejected gaseous matter in an equatorial disk 
around a rapidly rotating star (see, e.g., Porter \& Rivinius \cite{Porter}). 
In this model, line emission other than from hydrogen is typically much 
weaker. Neither the strong forbidden lines nor the arc-shaped circumstellar 
matter could be explained with this scenario.

\begin{table*}[t]
\centering
\caption{Position angles and separations relative to component HD~155448~A. 
         }
\setlength\tabcolsep{11.8pt}
\begin{tabular}{llcrrcc}
    \hline
    \hline
    \noalign{\smallskip}
    Observation  &  Source                   &  Component  
                 &  $\rho$                   &  $\theta$      
                 &  RA                       &  Dec           \\
        (year)   &                           &  
                 &  [$''$]                   &  [$^{\circ}$]  
                 &  [J2000.0]                &  [J2000.0]     \\
    \noalign{\smallskip}
    \hline
    \noalign{\smallskip}
    1928         &  DN2002                   &  A          
                 &  0                        &  0           
                 &                           &                \\
    1928         &  DN2002                   &  B          
                 &  2.9                      &  280           
                 &                           &                \\
    1928         &  DN2002                   &  C          
                 &  4.2                      &  16            
                 &                           &                \\
    1911         &  DN2002                   &  D          
                 &  9.9                      &  91            
                 &                           &                \\
    \noalign{\smallskip}
    \hline
    \noalign{\smallskip}
    2005         &  NACO                     &  A
                 &  ...                      &  ...
                 &  17:12:58.79   $\pm$ 0.01 s
                 &  $-$32:14:33.7 $\pm$ 0.1 s   \\
    2005         &  NACO                     &  B1         
                 &  2.84 $\pm$ 0.01          &  279.8 $\pm$ 0.3      
                 &  17:12:58.57   $\pm$ 0.02 s
                 &  $-$32:14:33.2 $\pm$ 0.1 s   \\
    2005         &  NACO                     &  B2         
                 &  2.26 $\pm$ 0.01          &  302.7 $\pm$ 0.3      
                 &  17:12:58.64   $\pm$ 0.01 s
                 &  $-$32:14:32.5 $\pm$ 0.1 s   \\
    2005         &  NACO                     &  C          
                 &  4.39 $\pm$ 0.01          &  16.6 $\pm$ 0.3      
                 &  17:12:58.89   $\pm$ 0.02 s
                 &  $-$32:14:29.5 $\pm$ 0.1 s   \\
    2005         &  NACO                     &  D          
                 &  9.43 $\pm$ 0.01          &  90.4 $\pm$ 0.3             
                 &  17:12:59.54   $\pm$ 0.02 s
                 &  $-$32:14:33.8 $\pm$ 0.1 s   \\
   \noalign{\smallskip}
   \hline
\end{tabular}
\\[2mm]
\flushleft
References: DN2002 (Dommanget \& Nys \cite{Dommanget02}), NACO (data in this 
work). The errors of the position angles and separations in the literature 
data are not known.
\label{table:positions}
\bigskip
\medskip
\end{table*}

The forbidden [\ion{O}{i}] emission line is observed in a 
wide range of astrophysical objects, and seen around both young and 
evolved stars. For Herbig Ae/Be systems, Acke et al.\ (\cite{Acke}) find 
a correlation between the strength of the [\ion{O}{i}] line at 6300\,\AA\ 
and the far-IR excess and show that stars with a flared disk geometry 
are more likely to exhibit the [\ion{O}{I}] line than self-shadowed disks. 
In these objects, the [\ion{O}{i}] line observed is centred at the star's 
velocity and is double peaked, therefore indicating that the emission arises 
from the circumstellar disk. The 
appearance of [\ion{O}{i}] together with [\ion{S}{ii}] in emission is commonly 
observed for accretion phenomena and outflows both in massive and 
low-mass stars (e.g., Hartmann \& Raymond \cite{Hartmann}; Hartigan 
et al.\ \cite{Hartigan}), while both lines are quite common in Herbig Ae/Be
systems. Vieira et al.\ (\cite{Vieira}) find the forbidden [\ion{O}{i}] 
lines (6300 and 6363\,\AA) and [\ion{S}{ii}] lines (6716 and 6731\,\AA) 
for 46\% of their 131 Herbig Ae/Be candidate stars, and find a higher
occurrence of those lines among B-type stars and those objects that have
stronger circumstellar emission. 

We see 
indications of a wind or outflow from HD~155448~C. The velocities of the
[\ion{S}{ii}] and [\ion{N}{ii}] lines, which are spatially extended in direction
towards the arc, point to a wind that is receding from us (cf.\ 
Sect.\,\ref{sect:emission-lines}) and that probably may interact with 
remnant matter of the star formation process located in the arc. Indeed, 
jets and collimated outflows from early B-type stars are observed for objects 
on the ZAMS (see, e.g., Shepherd \cite{Shepherd}). 

Why is the [\ion{O}{i}] emission not part of this receding wind? In a study 
of forbidden emission lines in T~Tauri and Herbig Ae/Be stars, Hamann et 
al.\ (\cite{Hamann}) found that the different lines are formed at distinct 
densities and 
temperatures in the circumstellar environment. While [\ion{O}{i}] needs a 
relatively high density, this is much lower for [\ion{N}{ii}] and even lower 
for [\ion{S}{ii}]. If an outflow exists from HD~155448~C, it implies a 
radial decrease in density with distance from the star. This agrees with our
observation, where [\ion{O}{i}] is found closer to the star than the other two 
forbidden lines (cf.\ Table~\ref{table:emission-lines}). These regions 
overlap to some extent, but each one is dominated by a different transition.

Multicomponent winds were also found in, e.g., \object{DG Tau} and 
\object{HL Tau} (Solf \& Boehm \cite{Solf}), where the low-velocity 
[\ion{O}{i}] 6300\,\AA\ components coincide with the star, while the 
high-velocity components [\ion{N}{ii}] and [\ion{S}{ii}] are distributed along 
the jet direction. No high-velocity components exist in HD~155448~C, 
but within the different velocities, this system might be a similar
scenario. The low wind velocities may be intrinsic or
caused by a projection effect. Some of the observed [\ion{O}{i}] emission  
might also originate in the disk around HD~155448~C, but the link to a 
wind origin relies on the fact that the [\ion{O}{i}] line is observed 
slightly extended to the east (see Fig.\,\ref{fig:pos-vel}).

{\it We note that the interpretation by means of a wind or outflow is not 
a unique solution of the HD~155448 system geometry, and discuss an 
alternative possibility at the end of Sect.\,\ref{sect:arc}.}

\subsection{The arc}
\label{sect:arc}

We could not find a scenario that perfectly explains the emission pattern 
observed around HD~155448~C. The silicates and PAHs represent remnant 
interstellar matter, the first one probably located in a disk around 
HD~155448~C, the latter in the more extended, asymmetric arc. There are 
two possible scenarios:

\begin{itemize}

\item{
A circumstellar disk or envelope exists around 
HD~155448~C, with the MIR flux dominated by emission from hot silicate 
grains. The arc to the northeast constitutes remnant matter from the stellar 
formation process. The 
PAHs in the arc are excited by the irradiation of all close B-type stars. 
The wind or outflow originating in HD~155448~C impacts the ISM and thus 
creates its arc-shaped form. 
}

\item{
In an alternative scenario we may speculate on the formation of 
another, so far undetected, late-type protostar in the HD~155448 system, 
located at the central part of the arc. As the stellar formation timescales 
are much larger for lower masses, typically 10$^7$ vs.\ 10$^4$ years, this 
protostar could still be embedded in its parental cloud. Indeed, in the 
mid-IR we see a knot in the central part of the arc, which is less prominent 
in the near-IR and not observed at optical wavelengths. The arc's shape may
have been caused by the interaction of the wind with the envelope of this 
hypothetical low-mass object. 
}

\end{itemize}

We favour the first scenario, because, although the other alternative is 
plausible, we were not able to provide more support for it with this work. 
Altogether, we suggest the following interpretation: there is an outflow from
HD~155448~C in the direction of the arc. The latter is composed of remnant 
material of the star-forming process and illuminated by the radiation of 
the stars. Its PAHs are partly ionised and resemble ISM material. The 
forbidden emission lines trace several wind components along the outflow,
with [\ion{O}{i}] centred on the star, where the outflow density is 
highest and the velocity component close to zero, while [\ion{N}{ii}] and 
[\ion{S}{ii}] are formed at greater distances from HD~155448~C, where the 
outflow density is lower and the wind velocity higher. 

An important limitation of the outflow interpretation is that only the 
receding (redshifted) outflow component is seen. Because in our observations 
we do not have evidence of material blocking the preceding (blueshifted) 
outflow component, we must conclude that the outflow in HD~155448~C has only 
the single component receding from us. There may be the following 
explanations.

\begin{itemize}

\item{
Lovelace et al.\ (\cite{Lovelace}) show that an intrinsically one-sided wind
can indeed occur for a rapidly rotating protostar with a magnetosphere in the 
propeller regime and with a magnetic field consisting both of a dipole and a 
quadrupole component. Whether such a configuration may apply to HD~155448~C 
is not known, since additional data would be needed for a detailed 
investigation of the magnetic field.
}

\item{
Alternatively, we may also interpret the redshifted forbidden line emission 
as originating in infalling remnant matter of the star-forming process. In 
this case the arc material will be situated in the foreground of HD~155448~C. 
}

\end{itemize}

With the data presented in this work, we cannot decide which of these
explanations is the more likely one. Spectra with both higher spectral and 
spatial resolution, ideally obtained with an integral field unit, are needed 
to better understand the physical interactions between HD~155448~C and its 
surrounding matter. To set constraints on a magnetic field of HD~155448~C, 
spectropolarimetric observations at a high spatial resolution are required. 
Given the rather faint magnitude of HD~155448~C, these follow-up observations 
will be ambitious.

\section{Conclusions}
\label{sect:conclusions}

We presented multi-wavelength photometry of the
\object{\mbox{HD 155448}} system, covering the optical and near- and mid-IR 
ranges, from the $B$ to the $Q$ bands. From the optical spectra we derived 
the spectral types of all five components. HD~155448~C shows strong emission 
lines, while circumstellar matter in an unusual, arc-shaped form is seen 
within 3\arcsec\ northeast of HD~155448~C. The main results of this
study follow.

\begin{enumerate}

\item{
Near-IR images obtained with ADONIS, SOFI, and NACO revealed that HD~155448~B 
is a binary. Including this discovery, the system now comprises five objects 
within $\sim$12\arcsec. Our distance estimate from spectral parallaxes 
indicates that all stars are at a similar distance of $\sim$2~kpc, but the 
error of 0.7~kpc is rather large, stemming from the fact that B~star 
luminosities vary strongly with one subclass uncertainty in the spectral 
classification.
}

\item{
From EFOSC2 optical spectroscopy of all five sources, we determined their 
spectral types as B1V (HD~155448~A), B6V (HD~155448~B1), B9V (HD~155448~B2),
B4Ve (C), and B8V (D).}

\item{
In a colour-magnitude diagram with stellar evolutionary tracks from Siess et 
al.\ (\cite{Siess}), the HD~155448 components B1, B2, C, and D are located
very close to the ZAMS (while the mass of HD~155448~A has no corresponding 
track in the Siess models). Earlier classifications of HD~155448 as a 
transition object from post-AGB to pre-PN are not consistent with this 
location.
}

\item{
HD~155448~C exhibits strong emission lines of H${\alpha}$ 
and [\ion{S}{ii}], as well as lines of [\ion{N}{ii}] and [\ion{O}{i}]. All 
emission lines in HD~155448~C are spatially extended in the eastern direction 
up to $\sim$4\arcsec and appear to have a similar velocity shift, except the 
[\ion{O}{i}] line, which is significantly less extended in spatial direction 
and close to the rest velocity of the star. 
}

\item{
No significant orbital motion in the HD~155448 system was observed, neither 
in the time span of our data sets between the year 2000 to 2005 nor when 
considering literature data from $\sim$80 years ago, though the uncertainties 
in the latter data did not permit an exact comparison.
}

\item{
Only HD~155448~C is detected in the mid-IR. From VISIR 10~$\mu$m spectroscopy, 
we see that the C component is surrounded by small, warm silicate grains, 
suggesting the presence of a circumstellar disk. In contrast, the emission 
features in the 
arc are entirely of PAHs. Their PAH band signature is more characteristic of 
ionised matter than of a circumstellar disk.
}

\item{
We find no scenario that could satisfactorily explain the shape and nature of 
the arc. A wind or outflow originating in HD~155448~C may impact the ISM, 
thereby creating its arc-shaped form. Alternatively, we may speculate about an
interaction of the wind with the envelope of a hypothetical low-mass object.
}

\item{
High spatial and spectral resolution observations of gas lines are needed to 
further study and understand this enigmatic system.
}

\end{enumerate}

\begin{acknowledgement}

We wish to express our thanks to the following people who helped us during
this project: Hermann B\"ohnhardt for sharing the ADONIS image of HD~155448
with us, Marc Audard and Linda Podio for very helpful discussions of the
interpretation of the forbidden lines in HD~155448~C, and the anonymous
referee for helpful suggestions. GM is partially 
supported by the Spanish Project AYA 2008-01727. AC acknowledges support 
from a Swiss National Science Foundation grant (PP002--110504). We thank the 
ESO observatory staff and telescope operators for support during the 
observations. This work made use of the SIMBAD astronomical database, 
operated at the CDS, Strasbourg, France.

\end{acknowledgement}

\clearpage

\begin{appendix}

\section{Details on the observations and data reduction}
\label{sect:observation}

\subsection{ADONIS NIR imaging}

The ESO Adaptive Optics Near Infrared System (ADONIS; Beuzit et 
al.\ \cite{Beuzit1}) was available on the 3.6\,m telescope at ESO La~Silla 
until September 2002. Typical Strehl ratios were around 0.1 in $J$ band and 
0.3 in $H$ band. ADONIS was coupled with the near-IR camera SHARPII+ 
which operates in the $J$ to $K$ band. Our data were acquired on June 8, 
2000 in the $H$ and $SK$ (short $K$) filter with total exposure times up to 
600 seconds. We attached a fully opaque coronographic mask in front of the 
Lyot (pupil) stop to reject the peak of the PSF for the brightest source, 
to increase the integration time and sensitivity in order to reveal 
fainter structures. Details on the coronograph and its performance are 
given in Beuzit et al.\ (\cite{Beuzit2}). For \mbox{\object{HD 155448}}, a 
mask of 1.4$''$ and a lens scale with 0.1$''$/pix were used. We co-added ten
exposures of 60 seconds in both filters. The mask's fixed position meant
that no jitter offsets were possible. Data cube clean-up, dark and 
flatfield correction, as well as sky subtraction were performed in 
standard ways. Figure~\ref{fig:slits} shows an ADONIS $H$ band exposure. No 
PSF subtraction was applied in this case.

Despite the high sensitivity of the ADONIS data, no reliable photometry could 
be obtained. There were problems both with the calibrator 
stars and with the camera's ADU conversion factor during this 
night, and weather conditions were not very stable.

\subsection{EFOSC2 optical imaging \& spectroscopy}

\subsubsection{Photometry}

Broadband photometry in the $B$, $V$, and $R$ filters was obtained on February 
26, 2006 with EFOSC2 (Buzzoni et al.\ \cite{Buzzoni}) at the 3.6m telescope in 
La Silla. These imaging data have a binning of  1 (i.e.\ 0.157$''$/pix). To 
better resolve the emission 
between the components, the data were re-sampled to 0.079$''$/pix. Eight 
frames of 1~sec integration time were co-added for each filter. Despite 
the short exposures, the A~component was saturated in the $V$ and $R$ 
filters. The B2 source is not resolved from B1 in the optical filters. 
Components A, B, and C have a flux error of 0.1~mag due to the problem 
in resolving their fluxes. The aperture photometry of source C does not 
include the arc-shaped emission region. Landolt standard fields (Landolt 
\cite{Landolt}) for zero points were observed just prior to the science data. 
In addition, we used SExtractor to extract the photometry and colour 
determination of all $\sim$2000 
sources in the EFOSC2 field (about 5.5\arcmin $\times$\,5.5\arcmin), in order 
to check whether more objects belong to the \object{HD 155448} system.
An $R$ band EFOSC2 image is shown in Fig.\,\ref{fig:efosc}.

\subsubsection{Spectroscopy}
\label{sect:obs-efosc-spec}

EFOSC2 spectra were obtained on March 1 and April 20, 2006, as well as on 
August 19, 2007 and August 12, 2008. The spectra obtained in 2006 were taken 
with grism~11 and a 1.0\arcsec\ slit, covering 3380--7520\,\AA\ at a 
dispersion of 2.04\,\AA/pix and a mean resolution $R$$\sim$400. Slit 
orientations were applied as shown in Fig.\,\ref{fig:slits}, to avoid a 
contamination by the neighbouring components as much as possible and to 
resolve the C source and the arc individually. In 2007, we re-observed 
the C component with a 1.0\arcsec\ slit and EFOSC2 grisms 5 (5200--9350\,\AA, 
2.06\,\AA/pix, $R$$\sim$300) and 18 (4700--6770\,\AA, 1.0\,\AA/pix, 
$R$$\sim$600), to obtain higher resolution and a redder spectral coverage. 
The slit orientation during the 2007 campaign was east-west, to cover only 
the C component (``Slit C only'' in Fig.\,\ref{fig:slits}). When the 
high-resolution holographic grisms 
19 (4441--5114\,\AA, 0.34\,\AA/pix, $R$$\sim$3000) and 20 (6047--7147\,\AA, 
0.55\,\AA/pix, $R$$\sim$2500) became available, we observed the system again
with a 0.5\arcsec\ slit in orientations as shown in Fig.\,\ref{fig:slits}
and Table~\ref{table:spec-obs}. Spectrophotometric standard stars were taken 
from the list of Hamuy (\cite{Hamuy92}, \cite{Hamuy94}) and observed with a 
wider 5.0\arcsec\ slit. Bias and flats were taken in a standard way. 
The wavelength calibration was done by correlating to a HeAr lamp spectrum.
The calibrated final spectra were averaged for consecutive exposures in each 
setting and the continuum normalised.

Further data processing was performed for the $R$$\sim$3000 spectra.
The 3$\sigma$ error in the wavelength calibration obtained is $\sim$4~km/s 
($\sigma$\,=\,1.2~km/s). To produce the position velocity diagrams discussed 
in Sect.\,\ref{sect:emission-lines}, the two-dimensional wavelength 
calibrated spectra were corrected for the curvature induced by the instrument 
along the spatial direction and the trace in the PSF along the dispersion 
direction. The curvature along the spatial direction was determined by fitting
a second degree polynomial to the centre position (determined by a Gaussian 
fit) of the bright sky [\ion{O}{i}] emission line near 6300\,\AA\ along the 
spatial axis. Then the spectra at each spatial position was shifted such that 
the centre of the sky [\ion{O}{i}] emission is always at the same pixel. The 
reference pixel was the pixel in which the [\ion{O}{i}] sky line crosses the 
peak of the star's continuum. The PSF trace was determined by fitting 
Gaussians to the PSF at several positions in the continuum and then fitting a 
second degree polynomial to the PSF centre positions. The spectra then were
shifted such that the centre of the PSF is always located at the same pixel.
The reference pixel is the median of the PSF centre positions.

\subsection{SOFI NIR imaging}

We acquired $JHK$ broadband images with SOFI (Moorwood et 
al.\ \cite{Moorwood}) at the La Silla NTT during May 7, 2004. SOFI's 
small field scale with 0.144$''$/pix was used. For each filter, ten exposures 
with random jitter box width of 20\arcsec\ were coadded. The total 
integration time in each passband amounts to 100 sec. Darks and dome flats
were taken in a standard way. Unfortunately, 
it was not possible to get separate photometry for the B1 and B2 components, 
owing to the low spatial resolution. The photometry of the C component 
is derived without the arc, placing the aperture -- as closely as possible 
-- only around the star. The derived magnitudes of source C might be 
slightly fainter than they really are, since with circular apertures it cannot
be avoided that the very closely located arc contributes partly to the 2~pixel 
thin sky annulus, thus resulting in a higher value for the sky. This 
uncertainty is considered in the corresponding magnitude errors. Zero points 
(ZPs) were derived from NICMOS photometric standard stars, which were observed 
two hours after HD~155448. We estimate a 
final photometric accuracy of 0.1~mag, as we do not accurately know the quality 
of sky transparency during these observations that were obtained from the ESO 
science archive. The estimate of 0.1~mag is derived from the maximal difference
in our ZPs to those of a photometric SOFI night.

\subsection{NACO NIR imaging}
\label{naco-imag}

Additional adaptive optics data were obtained with NAOS-CONICA (NACO; Lenzen 
et al.\ \cite{Lenzen}, Rousset et al.\ \cite{Rousset}) at the Paranal 
Observatory in service mode during ESO period 75 (i.e.\ between April and 
September 2005). We acquired images in narrow band filters (NB) at the 
central wavelengths 1.64~$\mu$m, 2.12~$\mu$m, 3.74~$\mu$m, and 4.05~$\mu$m, 
with a spatial resolution of 0.027$''$/pix and total integration times up to
435 sec. Dark, flatfield, and sky correction were made in the standard way. 
Ten individual exposures for each filter, with a jitter box width
of 10\arcsec, were shifted and co-added.

We think that some problems must have occurred during these observations,
because we observed strong flux losses when comparing the derived photometry to 
the expected values, as could already be recognised from the raw data. We 
immediately see that the exposures are not very sensitive, so that even the 
arc northeast of the C component is not seen in any of the NACO images. 
Nevertheless, we could derive useful relative photometry and astrometry for 
the narrow band filters NB\_1.64 and NB\_2.12.
Unfortunately, no standard star was observed for this service mode programme. 
Therefore, we first derived relative photometry and then scaled it to 
the 2MASS $H$ and $K$ band flux of the D component, since this component likely 
has the most accurate 2MASS flux, which is recognisable from the flags in the 
2MASS catalogues, while its larger distance to the neighbouring stars will 
avoid contamination. 
The resulting NACO photometry matches that of SOFI within the error ranges.

\subsection{VISIR MIR imaging \& spectroscopy}

\subsubsection{Photometry}
\label{sect:visir-imag}

We observed the \object{HD 155448} system between 8 and 19~$\mu$m with the 
mid-IR instrument VISIR (Lagage et al.\ \cite{Lagage}) at Paranal 
in service mode in period 75. In the $N$ band, dual imaging around the 
wavelength of PAH emission was performed with the VISIR filters PAH2 
(11.25~$\mu$m), and PAH2\_ref2 (11.88~$\mu$m) as the nearest continuum 
reference. The $Q$ band exposures were taken with the $Q2$ filter (18.72~$\mu$m), 
which -- for VISIR -- has the best $Q$ band sensitivity. The spatial resolution 
at both wavelengths is 0.075$''$/pix in the VISIR small field. All mid-IR 
exposures were performed with a standard chopping and nodding technique 
with a throw of 10$''$. Mid-IR standard stars for flux calibration were 
selected by the observatory staff from a list of mid-IR 
standard stars by Cohen (\cite{Cohen}), and were observed close in time and 
airmass to the science observations. 

Data cube chop and nod co-addition was done in the standard ways. Where 
necessary, we corrected for known artifacts (in the form of stripes) of the 
VISIR detector. The co-added exposure time for each PAH observation amounts to 
1085 sec. For the final image, we selected the five best out of six 
observations, spread over several months, resulting in 5425 sec (1.5 hours) 
total integration. The resulting PAH on/off images have the same exposure 
time and $S/N$. Both images were re-centred with an accuracy of 0.1 pix 
before subtraction. In the $Q$ band, each co-added observation sums
to 1987 sec. For the final image, the four best observations (out of five) 
were selected, resulting in a total exposure time of 7948 sec (2.2 hours).

The photometric accuracy is limited by the following uncertainties:
(1)~uncertainty in the aperture, which is negligible in our case with
errors smaller than 5~mJy, (2)~uncertainty in background subtraction: by 
co-addition of many exposures this influence got also strongly suppressed, 
and (3)~uncertainty in the flux calibration. The latter may be significant, 
because the atmospheric transmission often varied in the short time between 
observation of the science target and the consecutively observed standard 
star. Since several calibrated images are co-added, this effect may be 
neutralised, but nevertheless we estimate the photometric accuracy to 
be only within 10\%.

\subsubsection{Spectroscopy}
\label{sect:visir-spec}

Low-resolution $N$ band spectra ($R$$\sim$350 at 10~$\mu$m) were obtained 
between 8 and 13~$\mu$m with a 1.0$''$ slit. We aimed at having both the 
signal of the star and its arc-shaped circumstellar matter in the slit, 
so we rotated the slit to an orientation of 50$^{\circ}$ 
position angle (cf.\ Figs.\,\ref{fig:irs-slits} and~\ref{fig:mir_imag}). 
Our three VISIR spectral settings, centred at 8.8~$\mu$m, 
11.4~$\mu$m, and 12.4~$\mu$m, cover approximately the wavelength ranges
8.0--9.4~$\mu$m and 10.4--13.1~$\mu$m, after removing the outer range of 
each setting.

The spectroscopic frames were debiased and checked for a correct 
alignment of the spectrum with the x-axis of the frame. Two peaks are 
distinguishable in the spectral profile: the position of the star and
that of the circumstellar arc. We extracted the signal separately 
for both regions, with "a neutral zone" left in between, and coadded
the signal of corresponding nod positions. In addition, data 
from various observations between June and September 2005 were averaged. 
An initial ad-hoc wavelength calibration, derived from the central 
wavelength of each spectral setting and the spectral resolution per 
pixel, was refined by cross-correlating with atmospheric lines. The 
final error in wavelength accuracy $\delta \lambda / \lambda$ is 
$<10^{-3}$. 

Mid-IR standard stars for telluric correction were selected by the 
observatory staff and executed immediately before or after the science 
data. Flux was calibrated in the standard way by dividing
the target through the calibrator spectrum and multiplying the result
with a profile of the standard star taken from Cohen 
(\cite{Cohen}). Due to the varying atmospheric transparency in the time 
between observation of the science target and the standard star (typically 
less than 1 hour), the flux calibration is not accurate so we need to 
cross-correlate the spectral flux to our aperture photometry. Moreover, 
we have wavelength dependent slit losses: while we covered the entire 
stellar neighbourhood (silicate-dominated), the emission of the outer part 
of the circumstellar arc (PAH-dominated) was not included in the 1.0\arcsec\ 
slit. Therefore, we scaled the 11.4~$\mu$m silicate spectrum of the stellar 
vicinity to the aperture photometry of 0.44~Jy at 11.88~$\mu$m (cf.\ 
Table~\ref{table:single-flux-2}), motivated by the dominance of this spectrum 
by silicates. 
Similarly, we scaled the 11.4~$\mu$m spectrum of the PAH-dominated 
circumstellar arc to the photometry for the arc of 0.41~Jy at 
11.25~$\mu$m, which was obtained as the difference of the total flux and 
the flux in the vicinity of the star (cf.\ Table~\ref{table:single-flux-2}).
We corrected most of the slit losses in the PAH emitting wing by this
approach. Finally, the 12.4~$\mu$m settings, which overlap with the 11.4~$\mu$m 
ones, were scaled to match the flux of the latter. We note that this flux 
calibration is not a unique solution, but the best effort possible for this 
data. We estimate a possible deviation from the actual flux within 10\%. 
More difficult is the adjustment of the 8.8~$\mu$m settings, which do 
not overlap with the more longward settings and for which we do not have 
aperture photometry at shorter wavelengths. In the absence of other 
calibration options, the 8.8~$\mu$m settings are adjusted to match the 
longward spectra. The approximate error in flux calibration for the 
8.8~$\mu$m silicate spectrum is another 10\%, while the 
absolute intensity of the 8.6~$\mu$m PAH band remains uncertain.

\subsection{Spitzer MIR spectroscopy}

\object{HD 155448} was observed with the infrared spectrograph (IRS, Houck et 
al.\ \cite{Houck}) onboard the {\it Spitzer} Space Telescope on March 22, 2005 
(PID: 3470, PI: Jeroen Bouwman). The source was measured using Short Low 
(5.2--14.5~$\mu$m), Short High (9.9--19.5~$\mu$m), and Long High 
(18.7--37.2~$\mu$m) modules. The integration time was 30 seconds for the 
Short High module and six seconds for the other modules, respectively, and 
at least two cycles were used for redundancy. A high-accuracy PCRS peak-up 
was used to acquire the target in the spectrograph slit. The spectra are 
based on the {\tt droopres} and {\tt rsc} products, processed through the 
S18.7.0 version of the Spitzer data pipeline for the low- and high-resolution 
modules, respectively. For the details of the data reduction procedure we 
refer to Juh\'asz et al.\ (\cite{Juhasz2010}) and Acke et 
al.\ (\cite{Acke2010}). The brief description of the data reduction steps is 
as follows. For the Short Low spectrum the associated pairs of imaged spectra 
were subtracted in order to correct for the background emission, stray-light 
contamination and anomalous dark currents. For the high-resolution spectra 
the background has been removed by fitting a local continuum underneath the 
source profile. Pixels flagged by the data pipeline as being ``bad'' were 
replaced with a value interpolated from an 8-pixel perimeter surrounding the 
flagged pixel. In the case of the Short Low module, the spectra were extracted 
using a 6.0-pixel fixed-width aperture in the spatial dimension. For the 
high-resolution modules, spectral extraction was done by fitting the source 
profile with the known PSF in the spectral images. Low-level fringing at 
wavelengths $>$20~$\mu$m was removed using the {\tt irsfringe} package 
(Lahuis \& Boogert \cite{Lahuis}). The spectra were calibrated using a 
spectral response function derived from IRS spectra and MARCS stellar models 
for a suite of calibrators provided by the Spitzer Science Centre. To remove 
any effect of pointing offsets, we matched orders based on the point spread 
function of the IRS instrument, correcting for possible flux losses.

\end{appendix}

\end{document}